\documentclass[useAMS,usenatbib]{mn2e}

\usepackage{graphics,epsfig}
\usepackage{graphicx}
\usepackage{amssymb}
\usepackage{lscape}

\title[GMRT observations of $X$-shaped radio sources]
{GMRT observations of $X$-shaped radio sources}
\author[D. V. Lal and A. P. Rao]{Dharam Vir Lal\thanks{E-mail:
dharam@ncra.tifr.res.in} and 
A. Pramesh Rao \\
\\
National Centre for Radio Astrophysics (NCRA--TIFR),
Pune University campus, Ganeshkhind, Pune - 411 007, India.}
\begin{document}

\date{Accepted 1988 December 15. Received 1988 December 14; in original form 1988 October 11}

\pagerange{\pageref{firstpage}--\pageref{lastpage}} \pubyear{2002}

\maketitle

\label{firstpage}

\begin{abstract}
We present results from a study of $X$-shaped sources based on observations
using the Giant Metrewave Radio Telescope (GMRT).  These observations were
motivated by our low frequency study of 3C~223.1 (Lal \& Rao 2005),
an $X$-shaped radio source, which showed that the wings (or
low-surface-brightness jets) have flatter spectral indices than
the active lobes (or high-surface-brightness jets), a result not
easily explained by most models. We have now obtained GMRT data at 240 and
610~MHz for almost all the known $X$-shaped radio sources and have studied the
distribution of the spectral index across the sources.  While the radio
morphologies of all the sources at 240 and 610 MHz show the characteristic
$X$-shape, the spectral characteristics of the $X$-shaped radio sources,
seem to fall into three categories, namely, sources in which
(A) the wings have flatter spectral indices than the active lobes,
(B) the wings and the active lobes have comparable spectral indices, and
(C) the wings have steeper spectral indices than the active lobes.
We discuss the implications of the new observational results on the
various formation models that have been proposed for $X$-shaped sources.
\end{abstract}


\section{Introduction}
\label{introduction}

A peculiar and a very small subclass of extragalactic radio sources
called as the $X$-shaped or `winged' sources
are characterised by two low-surface-brightness
lobes (the `wings') oriented at an angle to the `active', or
high-surface-brightness radio lobes, giving the total source
an `$X$' shape. These two sets of lobes usually pass symmetrically through
the centre of the associated host galaxy.
\citet{MerrittEkers}
noted that majority of these sources
are of FR~II \citep{FanaroffRiley}
and rest are either FR~I or mixed.

$X$-shaped sources seem to reside in different types of host galaxies.
Nearly half of these sources reside in fairly elliptical host galaxies.
The hosts of some of these were reported to be slightly elongated and
in several prominent dust disks have been found \citep{Rottmann}.
The environments of $X$-shaped sources seem to be poor. Almost all
the sources are no part of rich clusters of groups.
\citet{Rottmann} reported, five (NGC~326, 3C~223.1, 4C 48.29,
B1059$+$169 3C~315 and 3C~403) of these sources to be located in
galaxy groups or clusters.
However the density of these galaxy groups
or clusters is low in all cases.  The X-ray result shows
that only one (NGC~326) is embedded in hot cluster gas, and this
is the only source having a close companion galaxy.

Radio observations of these $X$-shaped sources
show a high degree of polarisation (15\%--30\%) in the
wings and an apparent magnetic field structure parallel to the
edge of the source and along the length of the wings
\citep{Dennett}. Further high
frequency and high resolution radio polarisation
images showed field lines wrapping around the edges, as well as
complex internal structure (3C~223.1: \citealt{Blacketal};
3C~315: \citealt{Hogbom}; etc.).

Several authors have attempted to explain the unusual
structure in $X$-shaped sources. The first attempt was made by
\citet{Rees}, who suggested that the jet direction precesses
due to a realignment caused by the accretion of gas with
respect to the central black hole axis.
\citet{Dennett} discussed four possible scenarios
for the formation of such radio morphology: (1) backflow from the
active lobes into the wings \citep{LeahyWilliams,Capettietal};
(2) slow conical precession of the jet axis \citep{Parmaetal,Macketal};
(3) reorientation of the jet axis during which flow continues; and
(4) reorientation of the jet axis, but with the jet turned off or
at greatly reduced power during the change of direction.
\citet{MerrittEkers} suggested another possible scenario,
{\it i.e.} the reorientation of black holes's spin axis due to a minor
merger, leading to a sudden flip in the direction of any associated jet.
A variant of \citet{MerrittEkers} model was suggested by
\citet{Krishna}, where the sources with $Z$ morphology within
their $X$-shapes evolve along a $Z$--$X$ morphological sequence.
Presently, most of the observational results seem to prefer
possibilities 3, 4 of \citet{Dennett} or \citet{MerrittEkers},
and the key difference between these two
models is in terms of mechanism of reorientation;
former favoured the disc instability mechanism because of
little evidence for recent merger, while the latter
preferred the coalescence scenario. Nevertheless, in all these scenarios,
the wings are interpreted as relics of past
radio jets and the active lobes as the newer ones.

\begin{table*}
\caption{The observing log for all the observed $X$-shaped sources.}
\centering
\begin{tabular}{l|cccrcllcc}
\hline \\
       & RA & Dec & Observing & \multicolumn{2}{c}{Bandwidth} & Centre & \multicolumn{2}{c}{Calibrator} & t$_{\rm integration}$ \\
       & & & date & \multicolumn{2}{l}{Nominal ~~~Effective} & frequency & flux & phase & (on-source) \\
       &\multicolumn{2}{c}{(J2000)}& & \multicolumn{3}{c}{610~MHz~/~240~MHz} & &  & (hour) \\
\hline \\
4C~12.03 &00:09:52.6&$+$12:44:04.9&18 Dec 2003 & 16/8 & 13.5/6 & 606.44/237.69 & 3C~286 & 0054$-$035 & 4.97 \\
3C~52    &01:48:29.0&$+$53:32:35.4&12 Jan 2003 & 16/8 & 14.25/5 & 606.68/240.56 & 3C~286 & 0110$+$565 & 6.29 \\
      & & &  &      & &               &        & 0114$+$483 &       \\
3C~136.1 &05:16:03.1&$+$24:58:25.5&09 Jan 2003 & 16/8 & 14.25/6.75 & 606.68/240.06 & 3C~286 & 0521$+$166 & 5.65 \\
3C~192   &08:05:35.0&$+$24:09:50.0&10 Apr 2004 & 16/6 & 13.125/5.625 & 606.31/240.19 &  3C~48 & 0804$+$102 & 6.38 \\
B2~0828$+$32 &08:31:27.5&$+$32:19:26.4& 18 Dec 2003 & 16/8 & 13.125/5 & 606.31/236.44 & 3C~48 & 0909$+$428 & 6.94 \\
3C~223.1 &09:41:24.0&$+$39:44:41.9&18 Dec 2003 & 16/8  & 14.25/5 & 606.68/237.19 & 3C~286 & 0834$+$555 & 6.67 \\
4C~48.29 &10:20:53.6&$+$48:31:24.3&11 Apr 2004 & 16/6 & 13.875/5 & 606.44/240.12 & 3C~147 & 0834$+$555 & 6.42 \\
B1059$+$169&11:01:33.0&$+$16:43:52.0&08 Feb 2004 & 16/8 & 13.5/6 & 606.25/240.06 & 3C~286 & 1021$+$219 & 5.33 \\
3C~315   &15:13:40.1&$+$26:07:31.2&10 Jan 2003 & 8/6 & 6.375/5 & 610.13/237.19 & 3C~286 & 1506$+$375 & 5.89 \\
3C~403   &19:52:15.8&$+$02:30:24.5&18 Dec 2003 & 16/8 & 13.5/6.25 & 606.37/236.94 & 3C~48 & 1941$-$154 & 4.97 \\
3C~433   &21:23:44.5&$+$25:04:11.9&13 Jan 2003 & 8/6 & 6.375/5.625 & 610.13/240.75 & 3C~486 & 2225$-$049 & 5.05 \\
      & & &  &      & &               &        & 2052$+$365 &       \\
\\
\hline
\end{tabular}
\label{observ}
\end{table*}

\citet{LalRao2005} presented an unusual result for 3C~223.1 source,
{\it i.e.} the wings (or low-surface-brightness jets) have flatter
spectral indices with respect to the high-surface-brightness active lobes
and this result is not easily explained in most models of the
formation of $X$-shaped sources.
Although unusual, it is a valuable result which puts stringent
constraints on the formation models and nature of these sources.
This unusual result for 3C~223.1 provides the motivation of this paper,
{\it i.e.} systematic study of the sample of $X$-shaped sources
using Giant Metrewave Radio Telescope (GMRT).

Furthermore, in the same paper,
we presented an `alternative' formation scenario, which was not
addressed earlier, {\it i.e.} these sources consist of
two pairs of jets, which are associated with two unresolved AGNs.
We also presented some of the assumptions used
in the spectral ageing method for estimating the age,
in order to explain the unusual result.  Briefly, these are as follows:
(1) The injection spectral index is varying \citet{Palmaetal}.
(2)  In these sources, the low-surface-brightness,
wings are in the process of becoming new active jets.
Hence, it is not surprising that they have
flatter spectral index compared to the active lobes.
(3) Presence of some exotic reacceleration mechanism together with
standard Alfven waves and Fermi mechanisms.
(4) There is a gradient in magnetic field across the source,
together with a curved electron energy spectrum,
which would result in spectral indices being different
at distinct locations within the source \citep{Blundell}.

We here present the GMRT results from a systematic study of the sample of
$X$-shaped radio sources at 240 and 610~MHz.
In Section~\ref{observations}, we describe the observations of
the sample, data reduction (Section~\ref{data_reduction}) and
present the images derived from GMRT at 240 and 610~MHz along with
distribution of low frequency, 240--610~MHz spectra
across all these sources (Section~\ref{results}).
We also interpret our results,
combine our data with previously published data for
all of these sources, and discuss the statistical implications of these
results on the formation models (Section~\ref{discussion}).
We summarize the salient conclusions of our study in Section~\ref{conclusions}.

\section{Sample}
\label{sample}

The sample is drawn from the list of $X$-shaped sources mentioned in
\citet{MerrittEkers} compiled by \citet{LeahyParma}.
The sources have been selected solely on the basis of
their morphology.  3C~192 was classified, as an $X$-shaped source
\citet{Parmaetal}, but was not included
in the list of all such known sources \citep{MerrittEkers},
and we have included it in our sample.
However, it is important to note that the images on which the
selection was based have been obtained with various instruments
and with vast differences in sensitivity and resolution.
Therefore, the sample of these twelve sources is inhomogeneous and
in no sense a statistical complete sample.

\section{Observations}
\label{observations}

We adopted an observing strategy similar to our earlier
observations for 3C~223.1 \citep{LalRao2005}.
The 240 and 610~MHz feeds of GMRT
are coaxial feeds and therefore, simultaneous dual
frequency observations at these two frequencies were performed.
The primary beams are $\sim$108\arcmin and $\sim$43\arcmin at
240 and 610~MHz, respectively.
We made full synthesis observations of all the $X$-shaped sources
240 and 610 MHz, in the dual frequency mode, using the GMRT
in the standard
spectral line mode with a spectral resolution of 125 kHz.
These sources were observed in two
cycles (03DVL01: four of the eleven and 05DVL01: seven
of the eleven sources). Table~\ref{observ} gives
the details of the observations.
Since, NGC~326 is already observed using GMRT at low frequencies,
we did not perform new observations.

The GMRT has a hybrid configuration \citep{Swarupetal} with 14 of
its 30 antennas located in a central compact array with size $\sim$1.1~km
and the remaining antennas distributed in a roughly `Y' shaped
configuration, giving a maximum baseline length of $\sim$25~km.
The baselines obtained from antennas in the central
square are similar in length to the VLA~$D$-array,
while the baselines between the arm antennas are
comparable in length to the VLA~$B$-array.  Hence, a single
observation with the GMRT provides both,
it samples the UV plane adequately on the short baselines
as well as on the long baselines
and provides good angular resolution when mapping the detailed
source structure with reasonably good sensitivity.

\section{Data reduction}
\label{data_reduction}

The visibility data were converted to FITS and analyzed using standard AIPS.
The flux calibrators 3C~48, 3C~147 and 3C~286 were observed depending on the
availability either in the beginning and/or in the end as an
amplitude calibrator and to estimate and correct for
the bandpass shape. We used the flux density scale which is an extension
of the \citet{Baarsetal} scale to low frequencies,
using the coefficients in AIPS task `SETJY'.
The secondary phase calibrator were observed at intervals of 35~min.
The error in the estimated flux density,
both due to calibration and systematic, is $\lesssim$ 5\%.
The data suffered from scintillations and
intermittent radio frequency interference (RFI).
In addition to normal editing of the data, the
scintillations affected data and
channels affected due to RFI were identified and edited,
after which the central channels were averaged using AIPS task
`SPLAT' to reduce the data volume. To avoid bandwidth smearing,
effective band at 240 and 610~MHz
was reduced to 5 and 3 channels, respectively.

While imaging, 49 facets
spread across a $\sim$1.$^\circ8 \times1.^\circ$8 field were used at
240~MHz and 9 facets covering  slightly less
than a 0.$^\circ7\times0.^\circ7$ field, were used at 610~MHz to map
each of the two fields using AIPS task `IMAGR'.
In order to achieve high resolution images that are also sensitive
to extended structure, we have employed the
SDI CLEANing algorithm \citep{Steeretal}.
We used `uniform' weighting and the 3$-$D option for W~term
correction throughout our analysis.
The presence of a large number of point sources in the field
allowed us to do phase self-calibration to improve the image.
After 2--3 rounds of phase self-calibration, a final self-calibration
of both amplitude and phase was made to get the final image.
At each round of self-calibration, the image and the visibilities
were compared to check for the improvement in the source model.
The final maps were combined using AIPS task `FLATN' and corrected
for the primary beam of the GMRT antennas.

\section{Results}
\label{results}

The radio images shown in Figs.~\ref{full_syn_12_03} to~\ref{full_syn_433}
have nearly complete UV coverage, an angular resolution
$\sim$12$^{\prime\prime}$ and $\sim$5$^{\prime\prime}$ and the
rms~noise in the maps are in the range $\sim$1.0--6.9 and
$\sim$0.2--0.8 mJy~beam$^{-1}$ at 240 and 610~MHz, respectively.
The dynamic ranges in the maps are in the range 900--2000 and
1700--5000 at 240 and 610~MHz, respectively.
Consequently, in the vicinity of strong sources,
including the sources discussed here, the local noise was sometimes higher
than the noise in empty regions. The selection of contours shown in
figures is based on the rms noise in the immediate vicinity of the source,
with first contour level being 3--5 times this rms noise.
To make further comparisons of the morphology and flux densities,
the final calibrated UV data at 610~MHz
was mapped using UV~taper of 0--22~k$\lambda$,
which is similar to that of 240~MHz data and then restored
using the restoring beam corresponding to the 240~MHz map.
The full synthesis 610~MHz, 240~MHz contour maps and the 610~MHz matched
resolution contour map for all the observed sources
are shown in Figs.~\ref{full_syn_12_03} to \ref{full_syn_433}.
The sequence of maps is ordered in right ascension. An ellipse in a box
in the lower left-hand corner of each map shows the shape of the synthesized
beam (FWHM). All positions are given in J2000 coordinates.

\subsection{Radio morphology and low frequency radio spectra}

\begin{figure*}
\begin{center}
\begin{tabular}{ll}
\includegraphics[width=7.5cm]{MAPS/4C12_03_610_PUB.PS} &
\includegraphics[width=7.5cm]{MAPS/4C12_03_240_PUB.PS} \\ [-1.2cm]
\includegraphics[width=7.5cm]{MAPS/4C12_03_MATCH_PUB.PS} &
\includegraphics[width=7.5cm]{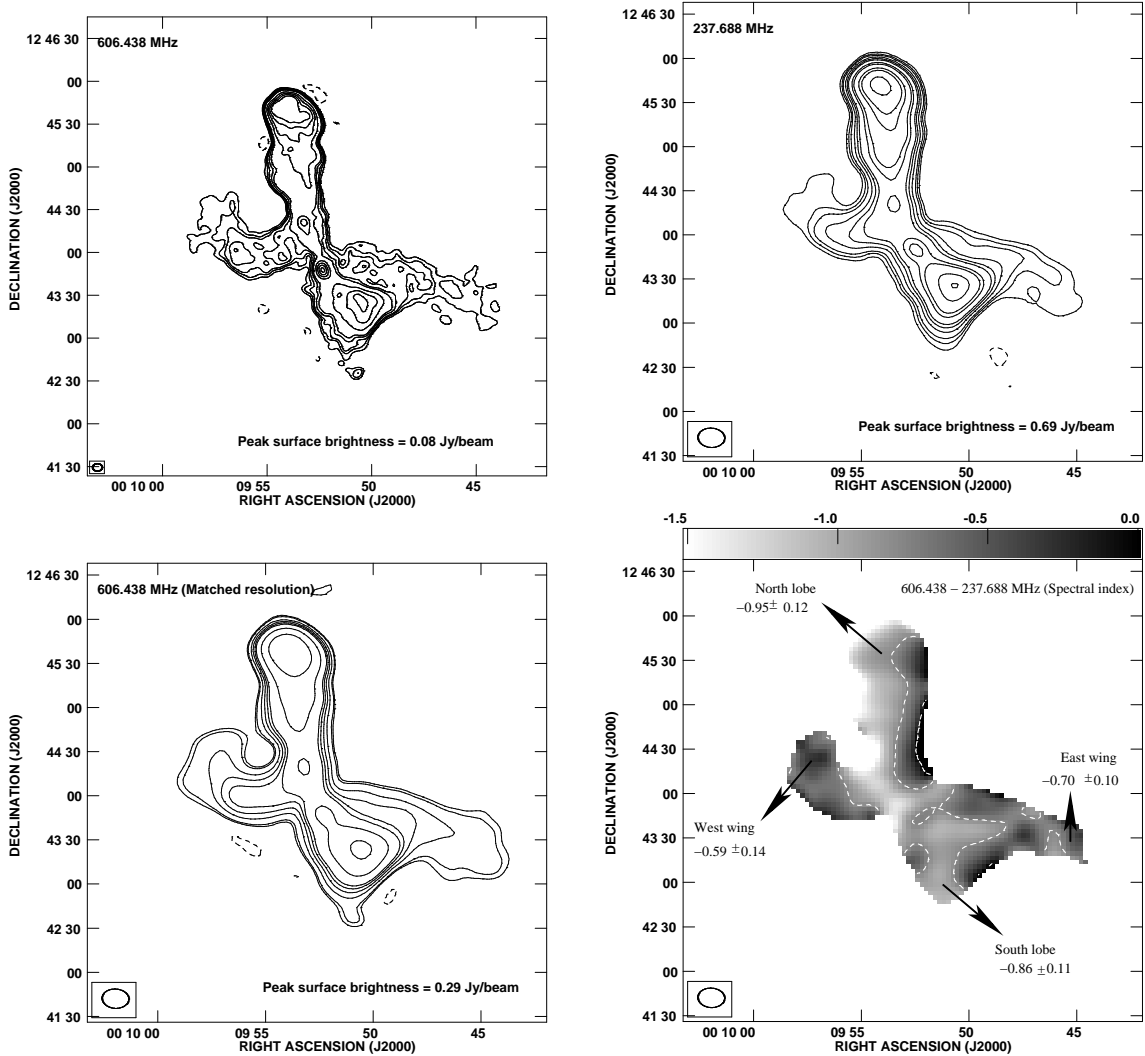} \\ [-0.8cm]
\end{tabular}
\end{center}
\caption{Upper: Full synthesis GMRT maps of 4C 12.03 at 610
(left panel) and 240~MHz (right panel).
The CLEAN beams for 610 and 240 MHz maps are
6$^{\prime\prime}$.7~$\times$~5$^{\prime\prime}$.5
at a P.A. of 89$^{\circ}$.8
and
18$^{\prime\prime}$.2~$\times$~13$^{\prime\prime}$.1
at a P.A. of 86$^{\circ}$.4, respectively;
and the contour levels in the two maps, respectively are
$-$2, 2, 3, 4, 6, 8, 16, 24, 32, 40 mJy~beam$^{-1}$
and
$-$20, 20, 30, 40, 60, 80, 100, 160, 200, 320, 400 mJy~beam$^{-1}$.
Lower left: The map of 4C 12.03 at 610~MHz
matched with the resolution of 240~MHz.
The contour levels are
$-$6, 6, 8, 16, 24, 32, 40, 80, 160 mJy~beam$^{-1}$.
Lower right: The distribution of the spectral index,
between 240 and 610 MHz, for the source.
The spectral index contours are at $-$0.8,~0.
The error-bars in the full synthesis maps found at a source free location
are $\sim$2.0 and $\sim$0.3~mJy~beam$^{-1}$ at 240 and 610~MHz, respectively.
}
\label{full_syn_12_03}
\end{figure*}

\begin{figure*}
\begin{center}
\begin{tabular}{ll}
\includegraphics[width=7.5cm]{MAPS/3C52_610_PUB.PS} &
\includegraphics[width=7.5cm]{MAPS/3C52_240_PUB.PS} \\ [-1.2cm]
\includegraphics[width=7.5cm]{MAPS/3C52_MATCH_PUB.PS} &
\includegraphics[width=7.5cm]{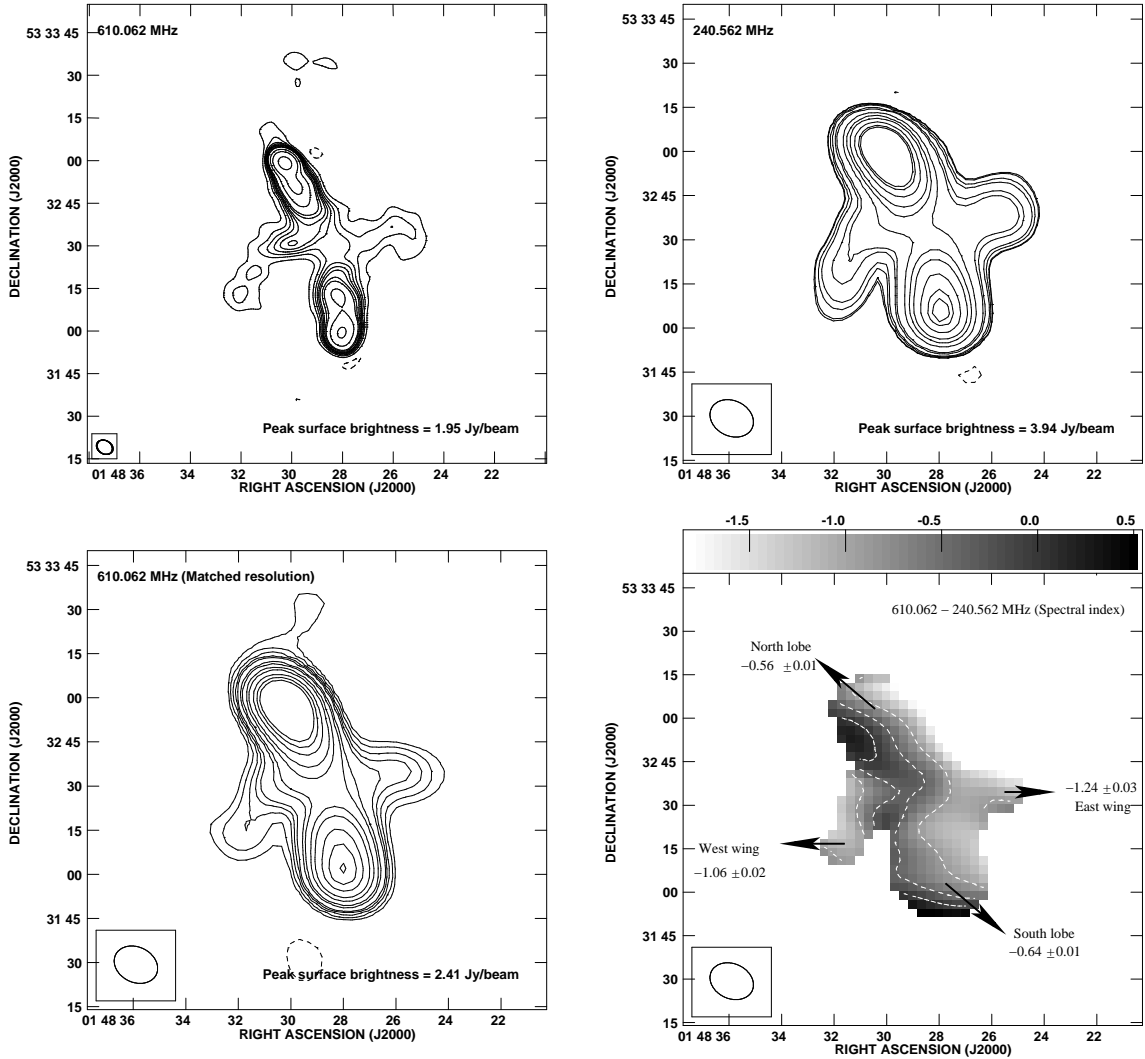} \\ [-0.8cm]
\end{tabular}
\end{center}
\caption{Upper: Full synthesis GMRT maps of 3C 52 at 610
(left panel) and 240~MHz (right panel).
The CLEAN beams for 610 and 240 MHz maps are
5$^{\prime\prime}$.9~$\times$~4$^{\prime\prime}$.7
at a P.A. of 57$^{\circ}$.5
and
15$^{\prime\prime}$.4~$\times$~12$^{\prime\prime}$.0
at a P.A. of 63$^{\circ}$.7,
respectively; and the contour levels in the two maps, respectively are
$-$8, 8, 20, 30, 40, 50, 60, 80, 100, 200 mJy~beam$^{-1}$
and
$-$50, 50, 60, 80, 100, 160, 200, 400, 600, 800, 1200, 1600, 1800 mJy~beam$^{-1}$.
Lower left: The map of 3C 52 at 610~MHz
matched with the resolution of 240~MHz.
The contour levels are
$-$24, 24, 40, 60, 80, 100, 160, 200, 300, 400, 600, 800 mJy~beam$^{-1}$.
Lower right: The distribution of the spectral index,
between 240 and 610 MHz, for the source.
The spectral index contours are at $-$0.9, $-$0.4,~0.
The error-bars in the full synthesis maps found at a source free location
are $\sim$1.4 and $\sim$0.3~mJy~beam$^{-1}$ at 240 and 610~MHz, respectively.
}
\label{full_syn_52}
\end{figure*}

\begin{figure*}
\begin{center}
\begin{tabular}{ll}
\includegraphics[width=7.5cm]{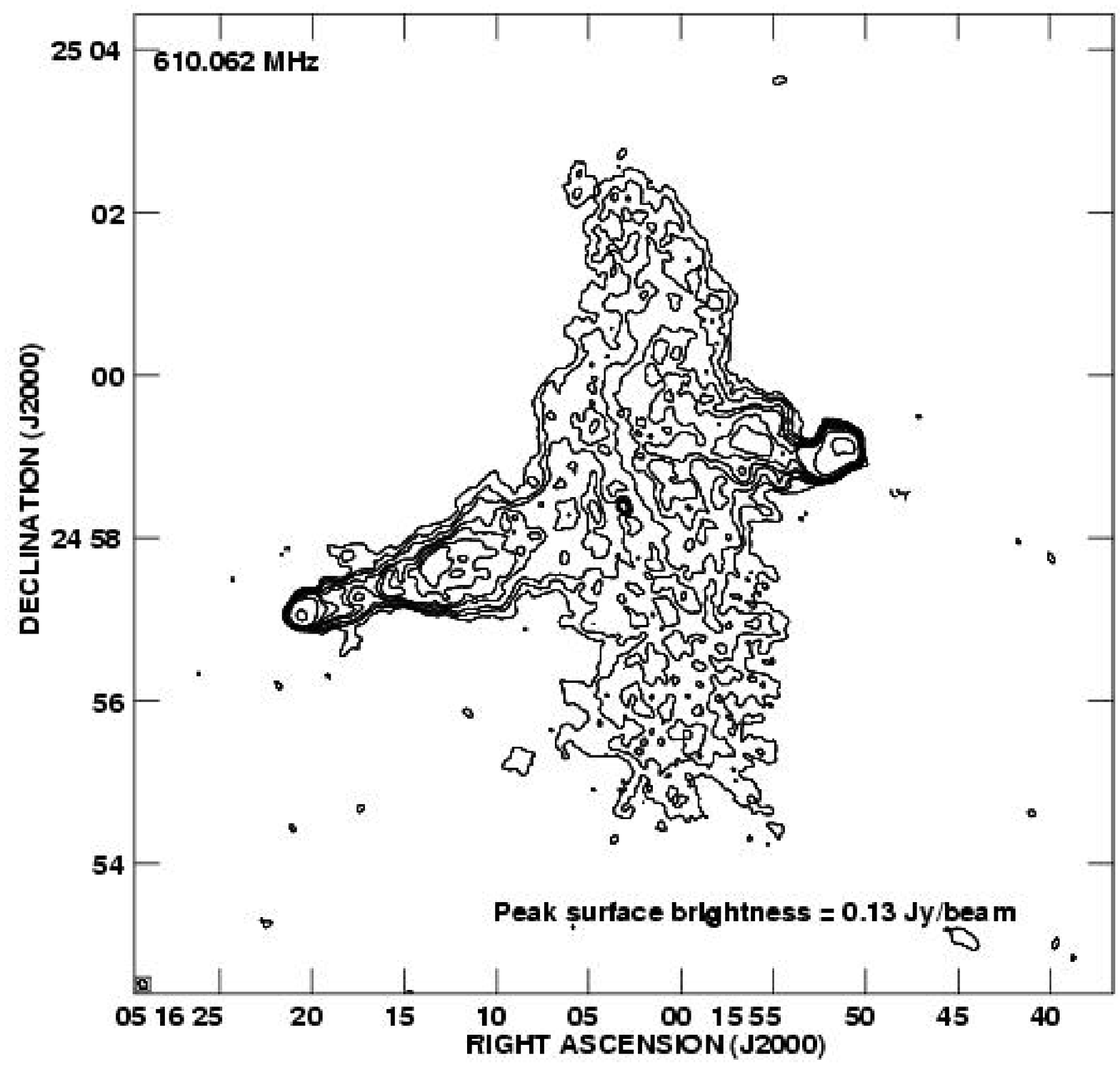} &
\includegraphics[width=7.5cm]{MAPS/3C136_1_240_PUB.PS} \\ [-1.2cm]
\includegraphics[width=7.5cm]{MAPS/3C136_1_MATCH_PUB.PS} &
\includegraphics[width=7.5cm]{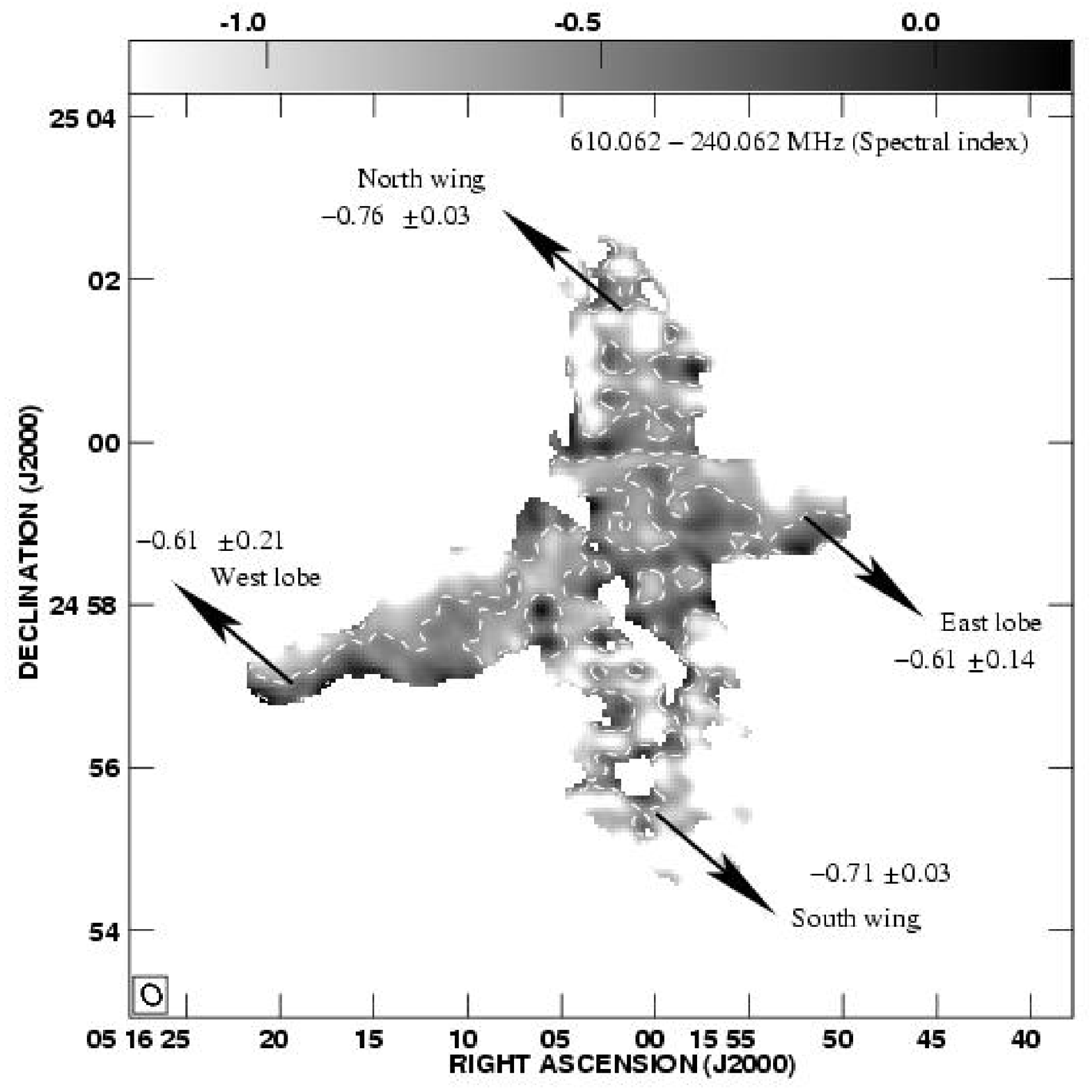} \\ [-0.8cm]
\end{tabular}
\end{center}
\caption{Upper: Full synthesis GMRT maps of 3C 136.1 at 610
(left panel) and 240~MHz (right panel).
The CLEAN beams for 610 and 240 MHz maps are
6$^{\prime\prime}$.7~$\times$~5$^{\prime\prime}$.1
at a P.A. of 28$^{\circ}$.2
and
15$^{\prime\prime}$.5~$\times$~12$^{\prime\prime}$.3
at a P.A. of 32$^{\circ}$.7,
respectively; and the contour levels in the two maps, respectively are
$-$1.6 1.6, 2.4, 3.2, 4.8, 6, 8, 10, 20, 80 mJy~beam$^{-1}$
and
$-$8, 8, 12, 20, 40, 60, 80, 100, 200, 400 mJy~beam$^{-1}$.
Lower left: The map of 3C 136.1 at 610~MHz
matched with the resolution of 240~MHz.
The contour levels are
$-$3, 3, 6, 8, 12, 20, 30, 40, 80, 200, 400 mJy~beam$^{-1}$.
Lower right: The distribution of the spectral index,
between 240 and 610 MHz, for the source.
The spectral index contours are at $-$0.6, 0.2.
The error-bars in the full synthesis maps found at a source free location
are $\sim$1.2 and $\sim$0.2~mJy~beam$^{-1}$ at 240 and 610~MHz, respectively.
The white patches in the spectral index map seen in the wings are
steep spectrum features and is due to slightly higher noise cut-off.
}
\label{full_syn_136_1}
\end{figure*}

The first high angular resolution, high sensitivity images
of $X$-shaped sources at the lowest frequencies of 240~MHz
(upper right-hand panels) and 610~MHz (upper left-hand panels)
are shown in Figs.~\ref{full_syn_12_03} to \ref{full_syn_433}.
In all the sources, the radio morphologies at 240 and 610 MHz show
well defined $X$-shape with a pair of active jets and a pair of wings,
that pass symmetrically through the position of the parent galaxy.
Table~\ref{flux_density} list the integrated flux densities
of all the sources along with previous measurements at other
frequencies and are plotted in Fig.~\ref{flux_reg}.
Our estimates at both frequencies, 240 and 610~MHz
agrees well with that of the measurements from other instruments.
We therefore believe that we have not lost any flux density in
our interferometric observations and there are no systematics
introduced in our analysis.

\begin{table*}
\centering
\caption{The total intensity for all the sources.
The total flux densities quoted are in Jy along with corresponding error-bars
(1$\sigma$). The 240 and 610 MHz are our GMRT measurements. The labels denote:
$^a$Large Cambridge interferometer \citep{Goweretal1967,Ryle};
$^b$The Molonglo reference catalogue of radio sources \citep{Largeetal1981};
$^c$Green Bank, Northern Sky Survey \citep{WhiteBecker};
$^d$\citet{GregoryCondon1991};
$^e$\citet{Kellermann};
$^f$\citet{Kuhretal1981};
$^g$Large Cambridge interferometer \citep{PilkingtonScott1965};
$^h$\citet{Ficarraetal};
$^i$VLA FIRST survey \citep{Beckeretal};
$^j$Green Bank, Northern Sky Survey \citep{WhiteBecker,BeckerWhite}.
}
\begin{tabular}{l|rrrrrrr}
\hline
\\
    & 178~MHz & 240~MHz & 408~MHz & 610~MHz & 1400~MHz & 2695~MHz & 4850~MHz \\
\\
\hline
\\
4C~12.03 & 7.6 $\pm$1.0$^a$ & 6.62 $\pm$0.12 & 4.45 $\pm$0.20$^b$ & 2.94 $\pm$0.05 & 2.01 $\pm$0.10$^c$ & & 0.54 $\pm$0.07$^d$ \\
3C 52 & 13.7 $\pm$1.1$^a$ & 11.71 $\pm$0.68 & & 6.44 $\pm$0.40 & 3.80 $\pm$0.19$^e$ & 2.30 $\pm$0.12$^e$ & 1.55 $\pm$0.17$^d$ \\
3C 136.1 & 14.0 $\pm$2.1$^e$ & 10.41 $\pm$0.05 & & 5.76 $\pm$0.03 & 2.90 $\pm$0.44$^e$ & 2.08 $\pm$0.10$^e$ & 0.56 $\pm$0.75$^d$ \\
3C~192 & 21.0 $\pm$3.2$^e$ & 20.51 $\pm$0.28 &11.03 $\pm$0.90$^f$ & 9.08 $\pm$0.12 & 4.80 $\pm$0.20$^f$ & 3.23 $\pm$0.16$^f$ & 2.68 $\pm$0.10$^f$ \\
B2 0828$+$32 & & 7.17 $\pm$0.05 & & 2.23 $\pm$0.02 & 2.07 $\pm$0.10$^c$ & & 0.44 $\pm$0.06$^d$ \\
3C~223.1 & 8.7 $\pm$1.1$^g$ & 8.33 $\pm$0.41 & 4.72 $\pm$0.10$^h$ & 3.56 $\pm$0.17 & 1.90 $\pm$0.28$^i$ & 1.23 $\pm$0.61$^e$ & 0.78 $\pm$0.11$^j$ \\
4C 48.29 & 4.5 $\pm$0.6$^a$ & 5.73 $\pm$0.04 & & 2.56 $\pm$0.03 & & & 0.35 $\pm$0.04$^d$ \\
B1059$+$169 & & 1.61 $\pm$0.02 & 1.02 $\pm$0.22$^b$ & 0.80 $\pm$0.01 & 0.62 $\pm$0.10$^c$ & & 0.21 $\pm$0.03$^d$ \\
3C 315 & 20.3 $\pm$1.7$^f$ & 39.83 $\pm$0.33 &10.62 $\pm$0.87$^f$ & 8.95 $\pm$0.09 & 4.10 $\pm$0.20$^f$ & 2.39 $\pm$0.05$^f$ & 1.28 $\pm$0.17$^d$ \\
3C 403 & 30.9 $\pm$4.0$^f$ & 17.62 $\pm$0.72 & 13.57 $\pm$0.59$^b$ & 9.81 $\pm$0.38 & 6.05 $\pm$0.19$^f$ & 3.65 $\pm$0.18$^f$ & 2.06 $\pm$0.10$^f$ \\
3C 433 & 60.4 $\pm$4.8$^f$ & 57.75 $\pm$3.98 & 29.26 $\pm$2.30$^f$ & 20.59 $\pm$1.21 & 12.40 $\pm$0.35$^f$ & 6.63 $\pm$0.33$^f$ & 4.05 $\pm$0.54$^d$ \\
\\
\hline
\end{tabular}
\label{flux_density}
\end{table*}

The observations and morphologies described
allow us to investigate in detail the spectral
index distributions of all sources.
The restored and matched maps at 240 and 610~MHz,
were used further for the spectral analysis for each of these sources.
We determine the spectral index distribution using
the standard direct method of determining the spectral
index between maps $S_{\nu_1}(x,y)$ and $S_{\nu_2}(x,y)$
at two frequencies ${\nu_1}$ and ${\nu_2}$, and is given by
the ratio of ${\rm log}~(S_{\nu_1} (x,y)/S_{\nu_2} (x,y))$
and ${\rm log}~(\nu_1/\nu_2)$.

The flux densities at 240 and 610~MHz plotted in
Fig.~\ref{flux_reg} are calculated using the images shown in
Figs.~\ref{full_syn_12_03} to \ref{full_syn_433} (upper right-hand
and lower left-hand panels),
which are matched to the same resolution,
and these values are tabulated in Tables~\ref{flux_density} and~\ref{x_age}.
The flux densities for the active lobes and the wings are
integrated over the region, which is at least four times the beam
size (a circular region of $\sim$5 pixels radius centered at the
position of tail of the arrows shown using AIPS task `IMEAN')
and above their 3$\sigma$ contour to reduce statistical errors.  In addition,
we have used conservative estimates of error-bars on the flux densities
at each location. These estimates were determined from the fluctuations in
the region being averaged and not from the noise at a source free location
using similar sized circular region, which being much smaller
(see Fig. captions).
These error-bars, both spectral indices and flux densities,
do not change significantly with increasing or decreasing the size of circle,
and they also do not change significantly by changing slightly the
position of circular region.
Furthermore, we have also examined the possibility that
(i) the different UV coverages,
(ii) the negative depression around the source and
(iii) the image misalignments
at 240 and 610 MHz could produce some systematic errors.
The former seems unlikely since the GMRT has good UV coverage, and
sources are only $\sim$3\arcmin--4\arcmin across and
are much smaller than the short baseline lengths,
$\sim$35\arcmin($\simeq$100 wavelengths) at 610 MHz and
$\sim$100\arcmin ($\simeq$35 wavelengths) at 240 MHz.
Nevertheless, we Fourier transform the
240~MHz CLEAN map, sampling it with the UV coverage of
610~MHz and re-imaging this visibility data set.
The resultant map showed no systematic differences from the original
240~MHz map and the rms difference in the two maps was less than 4\%,
corresponding to the rms error in the spectral index of $\lesssim$0.05.
Furthermore, 240~MHz maps of two sources
(Fig.~\ref{full_syn_223_1}: 3C~223.1 and Fig.~\ref{full_syn_433}: 3C~433)
show marginal evidence that these images contain a
negative depression around them.
Negative depression/bowl is seen in a synthesis image, either due to inadequate
UV coverage at short baseline lengths or due to inadequate CLEANing.
The former is unlikely as explained above, examining the
latter--While imaging, we did not provide zero-spacing flux density, but we
have done deep CLEANing, so as not to make any deconvolution errors.
Comparisons of expected integrated flux density, total CLEANed
flux density and flux density measured by short baseline lengths
suggest no discrepancy beyond 5\%.
In any case, we quantify the errors that would be introduced due to
possible negative depression for these two sources below.
Finally, in each case, we not only registered the target source,
the positions of at least five field sources around the $X$-shaped
source, which were common in the two maps were also registered,
and the two images alignment is better than 15\%.
Therefore, in order to understand the nature of $X$-shaped sources,
we take into account our careful analyses and estimation of the
possible systematic errors.

Analysis of the spectrum, shown in Figs.~\ref{full_syn_12_03} to
\ref{full_syn_433} (lower right-hand panels), in different regions of
each of these sources show remarkable variation across them.
The lighter regions represent the relatively steep spectrum
as compared to the darker regions which represent flat spectrum.
Although the full range of spectral index is large, we have shown
only small range for clarity in each case.
We now describe the radio structure of $X$-shaped radio sources
measured by the GMRT, along with the best power-law fit
($S_\nu \propto \nu^\alpha$) for several regions across the source.
Here, we~also include the description of 3C~223.1 \citep{LalRao2005} and
NGC~326 \citep{Murgiaetal}, published in the literature.

\subsubsection*{NGC~326 (z = 0.047)}

The galaxy is the brightest member of the Zwicky cluster
0056.9$+$2636 \citep{ZwickyKowal1968}.
The host is a dumbbell shaped galaxy with clearly
separated nuclei \citep{Wirth1982}.

NGC~326 was the first $X$-shaped radio source discovered
\citep{Ekers1978}. At 1.4~GHz, the lobes are slightly resolved and
the most prominent components are the two wings.
The lobes are asymmetric in total emission, extent and
distance from the core. The southern lobe has an ellipsoidal shape,
while the northern lobe is more elongated and wider \citep{Murgiaetal}.
The wings bend and
extend away from the lobe axis by $\sim$2\arcsec. Furthermore, the
overall $Z$-shape symmetry is broken by the low-surface-brightness
plume located just above the end of the east wing \citep{Worralletal}.

In the east wing there is monotonic steepening of the radio
spectrum from the south lobe to the end of the wing;
the spectral index distributions between 1.4 and 4.8~GHz
and between 4.8 and 8.5~GHz
decrease from $-$0.6 and $-$0.7 up to $-$1.3 and $-$1.9, respectively.
Whereas in the west wing,
the spectral index distributions between 1.4 and 4.8~GHz
and between 4.8 and 8.5~GHz decrease, respectively
from $-$0.6 and $-$0.7 up to $-$1.3 and $-$1.5 \citep{Murgiaetal}.
Similarly, in the south lobe
the spectral index distribution between 1.4 and 4.8~GHz
is roughly constant around a value of 0.6 and
the spectral index distribution between 4.8 and 8.5~GHz
decreases from $-$0.7 to $-$1.3.  Whereas in the north lobe
the spectral index distributions between 1.4 and 4.8 GHz
and between 4.8 and 8.5~GHz
increase, respectively from $-$0.75 and $-$1.35 at the lobe head, to $-$0.65
and $-$0.8 and becomes $-$1.0 and $-$1.6 in proximity of the core
\citep{Murgiaetal}.
Briefly, the active lobes have flatter
1.4--8.5~GHz spectral index as compared to the wings and
the 0.325--1.4~GHz spectra
also shows similar behaviour (private communication).

\subsubsection*{4C~12.03 (z = 0.110)}

4C~12.03 is associated with an elliptical host galaxy \citep{Heckman1994}
and has been classified as a low emission line radio galaxy
\citep{Laing1983}.  Morphologically, 4C~12.03 seems to lie at the
FR~I/FR~II division, whereas the radio luminosity suggests it to be
a FR~I source.

Fig.~\ref{full_syn_12_03} at 240 and 610~MHz
shows symmetrical structure and extent in the two
matched resolution radio maps, as is usually the case for radio
galaxies. The northern jet leading to north hot~spot possibly
consists of the active axis and the other axis, {\it i.e.} the east-west
axis consists of the wings.
The source has an angular extent of 3.8\arcmin $\times$ 4.0\arcmin
in both, 240 and 610~MHz maps.

The high frequency spectral index distribution between
1.5 and 10.45~GHz at a resolution of 69\arcsec $\times$ 69\arcsec
shows marginal steepening from the active lobes ($\alpha$ $\simeq$ $-$0.8)
to the wings ($\alpha$ $\simeq$ $-$1.0) \citep{Rottmann}.
The low frequency fitted spectra have
$-$0.59 $> \alpha >$ $-$0.95 for all regions across the source.
Contrary to the high frequency spectral results,
the low frequency result shows definite evidence for steeper spectra
in the active lobes than in the wings, and the east and west wings have
spectral indices,
$-$0.70 $\pm$0.10 and $-$0.59 $\pm$0.14, respectively,
whereas the north and south active lobes have
$-$0.95 $\pm$0.12 and $-$0.86 $\pm$0.11, respectively.

\begin{figure*}
\begin{center}
\begin{tabular}{ll}
\includegraphics[width=7.5cm]{MAPS/3C192_610_PUB.PS} &
\includegraphics[width=7.5cm]{MAPS/3C192_240_PUB.PS} \\ [-1.2cm]
\includegraphics[width=7.5cm]{MAPS/3C192_MATCH_PUB.PS} &
\includegraphics[width=7.5cm]{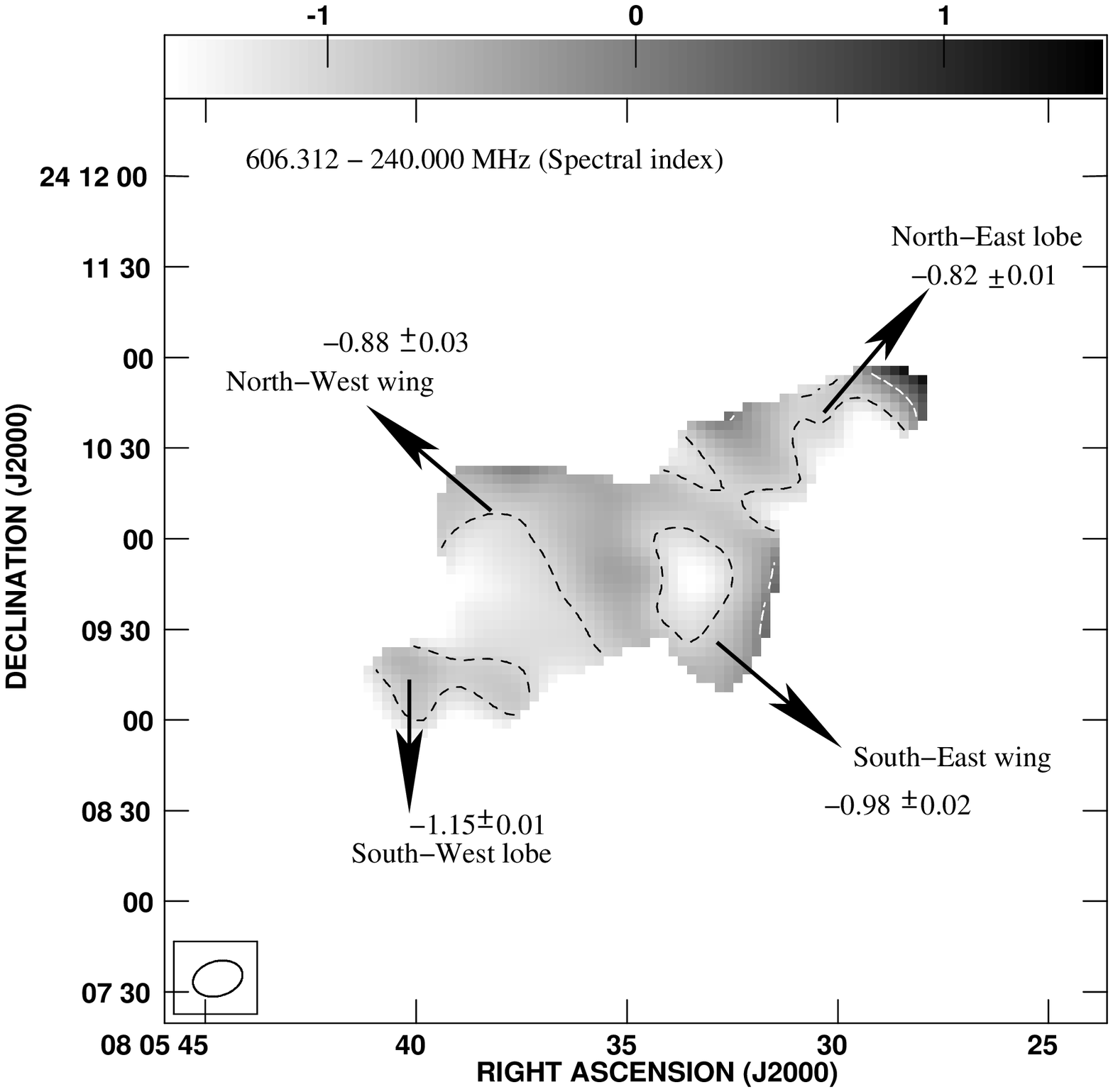} \\ [-0.8cm]
\end{tabular}
\end{center}
\caption{Upper: Full synthesis GMRT maps of 3C 192 at 610
(left panel) and 240~MHz (right panel).
The CLEAN beams for 610 and 240 MHz maps are
6$^{\prime\prime}$.0~$\times$~4$^{\prime\prime}$.5
at a P.A. of $-$88$^{\circ}$.1
and
16$^{\prime\prime}$.5~$\times$~11$^{\prime\prime}$.5
at a P.A. of $-$73$^{\circ}$.2,
respectively; and the contour levels in the two maps, respectively are
$-$4, 4, 6, 8, 10, 16, 24, 40, 60, 80 100 mJy~beam$^{-1}$
and
$-$50, 50, 60, 80, 100, 200, 400, 600, 800 1000 mJy~beam$^{-1}$.
Lower left: The map of 3C 192 at 610~MHz
matched with the resolution of 240~MHz.
The contour levels are
$-$20, 20, 40, 60, 80, 100, 200, 300, 400, 600, 800, 1000 mJy~beam$^{-1}$.
Lower right: The distribution of the spectral index,
between 240 and 610 MHz, for the source.
The spectral index contours are at $-$1,~0.
The error-bars in the full synthesis maps found at a source free location
are $\sim$2.4 and $\sim$0.5~mJy~beam$^{-1}$ at 240 and 610~MHz, respectively.
}
\label{full_syn_192}
\end{figure*}

\begin{figure*}
\begin{center}
\begin{tabular}{ll}
\includegraphics[width=7.5cm]{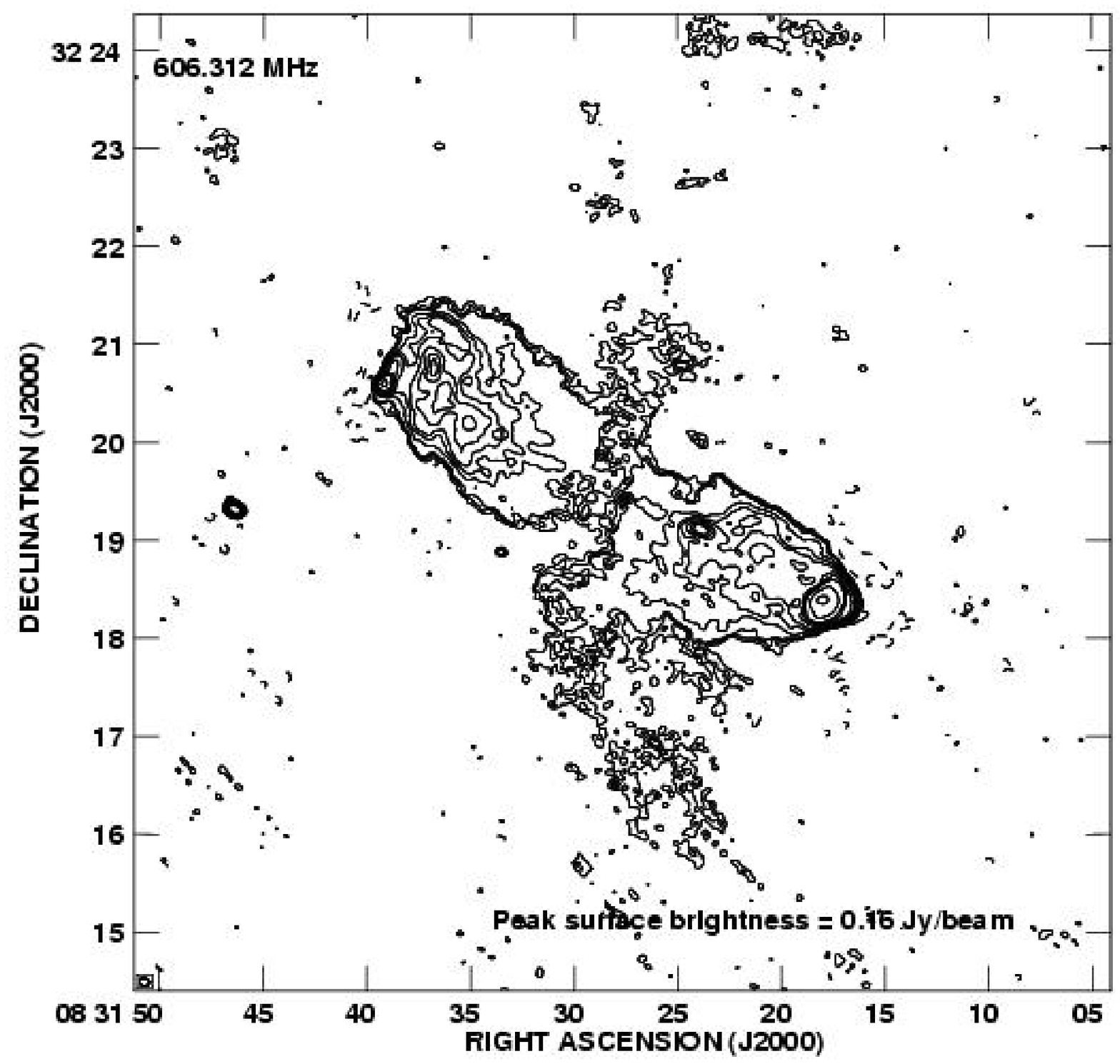} &
\includegraphics[width=7.5cm]{MAPS/B2_0828_32_240_PUB.PS} \\ [-1.2cm]
\includegraphics[width=7.5cm]{MAPS/B2_0828_32_MATCH_PUB.PS} &
\includegraphics[width=7.5cm]{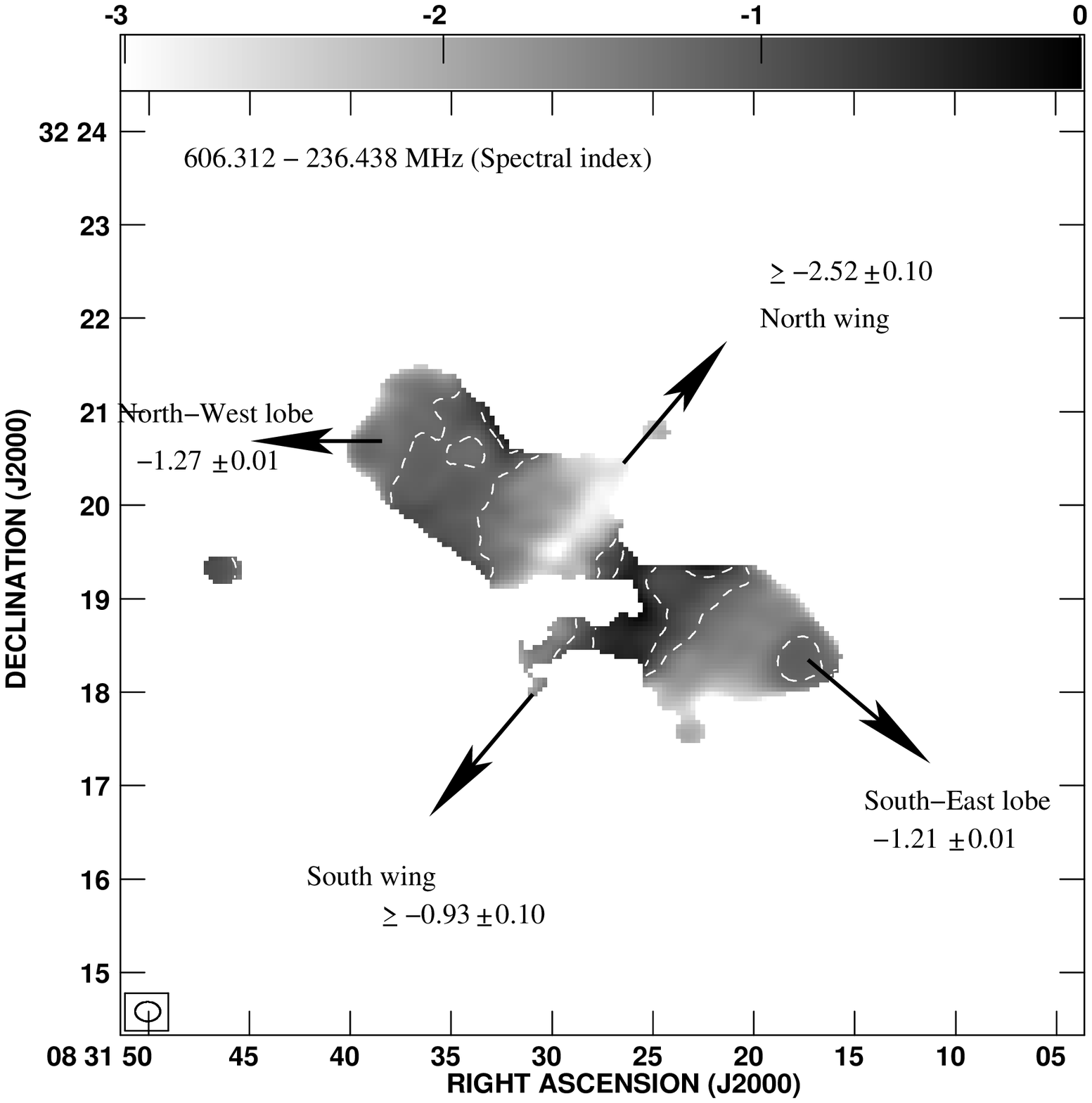} \\ [-0.8cm]
\end{tabular}
\end{center}
\caption{Upper: Full synthesis GMRT maps of B2~0828$+$28 at 610
(left panel) and 240~MHz (right panel).
The CLEAN beams for 610 and 240 MHz maps are
6$^{\prime\prime}$.0~$\times$~4$^{\prime\prime}$.8
at a P.A. of 81$^{\circ}$.9
and
15$^{\prime\prime}$.8~$\times$~12$^{\prime\prime}$.5
at a P.A. of 89$^{\circ}$.2,
respectively; and the contour levels in the two maps, respectively are
$-$1, 1, 2, 3, 4, 6, 8, 10, 20, 40 mJy~beam$^{-1}$
and
$-$14, 14, 20, 30, 40, 50, 80, 100, 120, 200, 400 mJy~beam$^{-1}$.
Lower left: The map of B2~0828$+$28 at 610~MHz
matched with the resolution of 240~MHz.
The contour levels are
$-$1, 1, 2, 4, 8, 10, 20, 30, 40, 60, 80, 100 mJy~beam$^{-1}$.
Lower right: The distribution of the spectral index,
between 240 and 610 MHz, for the source.
The spectral index contours are at $-$1.2, $-$0.7,~0.
The error-bars in the full synthesis maps found at a source free location
are $\sim$2.7 and $\sim$0.2~mJy~beam$^{-1}$ at 240 and 610~MHz, respectively.
}
\label{full_syn_b2}
\end{figure*}

\subsubsection*{3C~52 (z = 0.285)}

3C~52 is the most distant $X$-shaped source.
{\it HST} imaging by \citet{deKoff} showed the galaxy to be elongated
along the north-south axis and it shows a pronounced dust disk.

Fig.~\ref{full_syn_52} shows the radio images at 240 and 610~MHz.
The core is undetected in low resolution radio maps, but closer
inspection of radio contours in the 610~MHz map shows presence
of possible core at the position of parent galaxy.
Historically, this source is an example of a mirror symmetric
distortion, but it is now described by rotational symmetry
\citep{LeahyWilliams}.

High frequency spectral index constructed at a resolution of
5\arcsec.0 $\times$ 4\arcsec.8 using images at 1.4, 1.7 and 2.7~GHz
show indication for a spectral steepening towards the wings
\citep{Rottmann}.  Author also finds prominent spectral steepening
from the south lobe towards the west wing as compared to
mild spectral steepening from the north lobe towards the east wing.
Similar spectral steepening between 1.4 and 5.0~GHz
from south lobe towards the east wing and from north lobe
towards the west wing was also found by \citet{AlexanderLeahy1987}
The low frequency fitted spectra have
$-$0.56 $> \alpha >$ $-$1.24 for all regions across the source.
The source shows evidence for flatter spectra in the
active lobes than in the wings. The east and west wings have
spectral indices,
$-$1.24 $\pm$0.03 and $-$1.06 $\pm$0.02, respectively,
whereas the north and south
active lobes have $-$0.56 $\pm$0.01 and $-$0.64 $\pm$0.01, respectively.

\subsubsection*{3C~136.1 (z = 0.064)}

It is a low galactic latitude object in the sample.
{\it HST} image of the host galaxy \citep{Martel1999}
shows flattened and warped host and bears no resemblance with an elliptical
galaxy.  The host galaxy possibly has two/three nuclei and
seems to be showing irregular, disrupted, possibly by tidal forces,
signs of on-going merger.

Fig.~\ref{full_syn_136_1} shows the radio images at 240 and 610~MHz,
maps show similar extent and morphology along
both axes, similar to its radio map at high frequency 
\citep{LeahyWilliams}. Therefore their suggestion
that some of the large-scale structure may be missing from their map is
unlikely. The core is clearly detected in 610~MHz maps and is marginally
detected in 240~MHz map.

The spectral index maps of 3C~136.1 using radio maps
at 1.37, 4.85 and 10.45~GHz at
resolutions of 69\arcsec $\times$ 69\arcsec and 147\arcsec $\times$ 147\arcsec
show a spectral gradient from the active lobes
($-$0.50 $> \alpha >$ $-$0.65) towards the wings \citep{Rottmann}.
The spectral gradient is more pronounced along the east lobe
($\alpha$ = $-$0.6) to the south wing ($\alpha$ = $-$1.0)
as compared to the spectral gradient along the west lobe
($\alpha$ = $-$0.7) to the north wing ($\alpha$ = $-$0.85).
\citet{AlexanderLeahy1987} also found spectral steepening between
1.4 and 5.0~GHz from west lobe towards the core and from east lobe
towards the north wing using VLA and Cambridge 5~km telescope.
The low frequency fitted spectra have
$-$0.61 $> \alpha >$ $-$0.76 for all regions across the source.
Similar to the high frequency result,
our result also shows evidence for
steeper spectra in the wings than in the active lobes and
the north and south wings have spectral indices,
$-$0.76 $\pm$0.03 and $-$0.71 $\pm$0.03, respectively,
whereas the east and west active lobes have
$-$0.61 $\pm$0.14 and $-$0.61 $\pm$0.21, respectively.

\begin{figure*}
\begin{center}
\begin{tabular}{ll}
\includegraphics[width=7.5cm]{MAPS/3C223_1_610_PUB.PS} &
\includegraphics[width=7.5cm]{MAPS/3C223_1_240_PUB.PS} \\ [-1.2cm]
\includegraphics[width=7.5cm]{MAPS/3C223_1_MATCH_PUB.PS} &
\includegraphics[width=7.5cm]{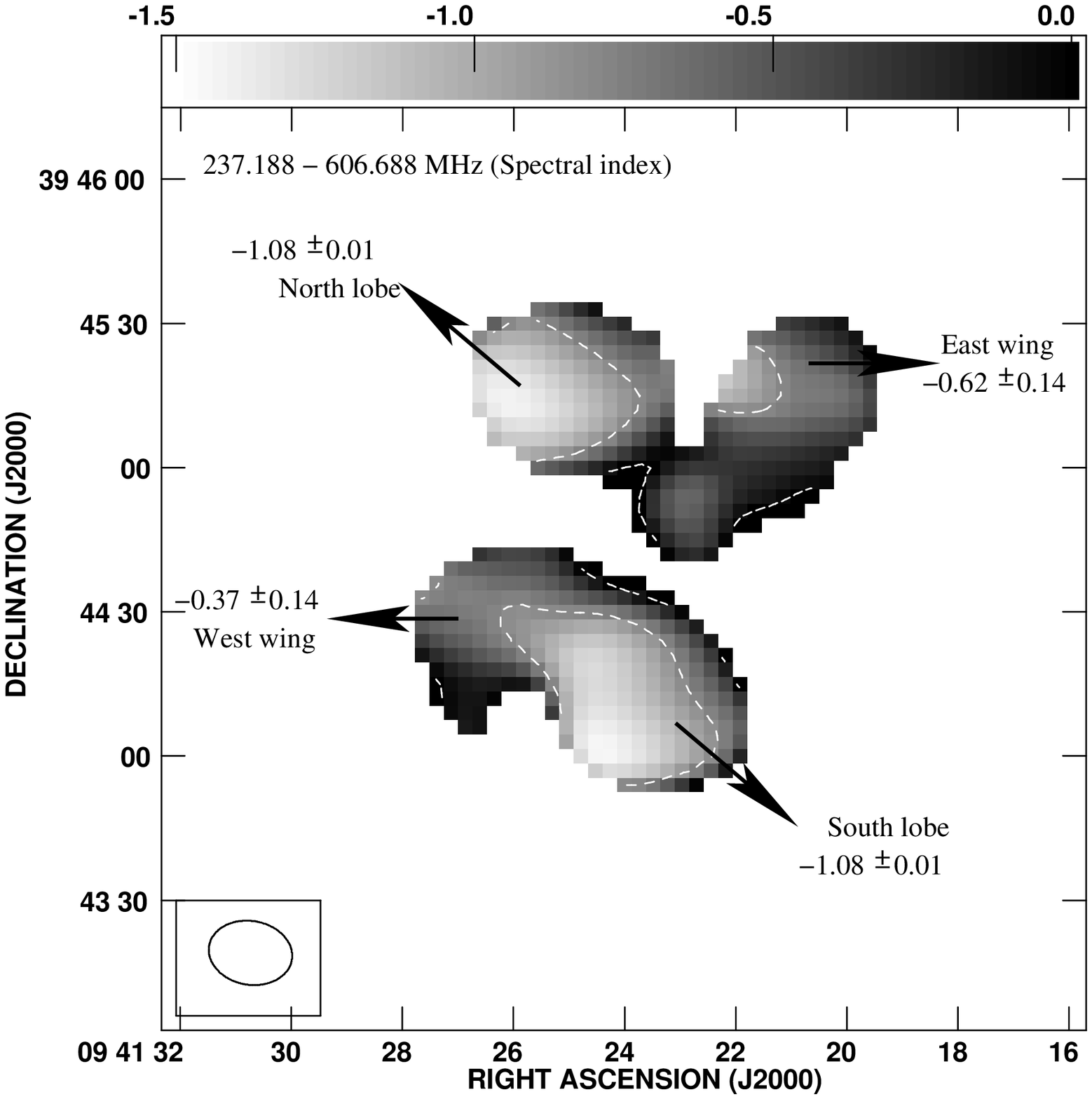} \\ [-0.8cm]
\end{tabular}
\end{center}
\caption{Upper: Full synthesis GMRT maps of 3C 223.1 at 610
(left panel) and 240~MHz (right panel).
The CLEAN beams for 610 and 240 MHz maps are
6$^{\prime\prime}$.1~$\times$~4$^{\prime\prime}$.8
at a P.A. of 66$^{\circ}$.8
and
17$^{\prime\prime}$.4~$\times$~13$^{\prime\prime}$.3
at a P.A. of 80$^{\circ}$.3,
respectively; and the contour levels in the two maps, respectively are
$-$2, 2, 4, 6, 8, 10, 12, 20, 40, 80, 100, 200, 300, 400 mJy~beam$^{-1}$
and
$-$32, 32, 60, 80, 100, 200, 400, 800, 1000, 4000 mJy~beam$^{-1}$.
Lower left: The map of 3C 223.1 at 610~MHz
matched with the resolution of 240~MHz.
The contour levels are
$-$10, 10, 20, 40, 60, 80, 100, 200, 400, 800, 1000 mJy~beam$^{-1}$.
Lower right: The distribution of the spectral index,
between 240 and 610 MHz, for the source.
The spectral index contours are at $-$0.8, 0.
The error-bars in the full synthesis maps found at a source free location
are $\sim$3.0 and $\sim$0.4~mJy~beam$^{-1}$ at 240 and 610~MHz, respectively.
}
\label{full_syn_223_1}
\end{figure*}

\begin{figure*}
\begin{center}
\begin{tabular}{ll}
\includegraphics[width=7.5cm]{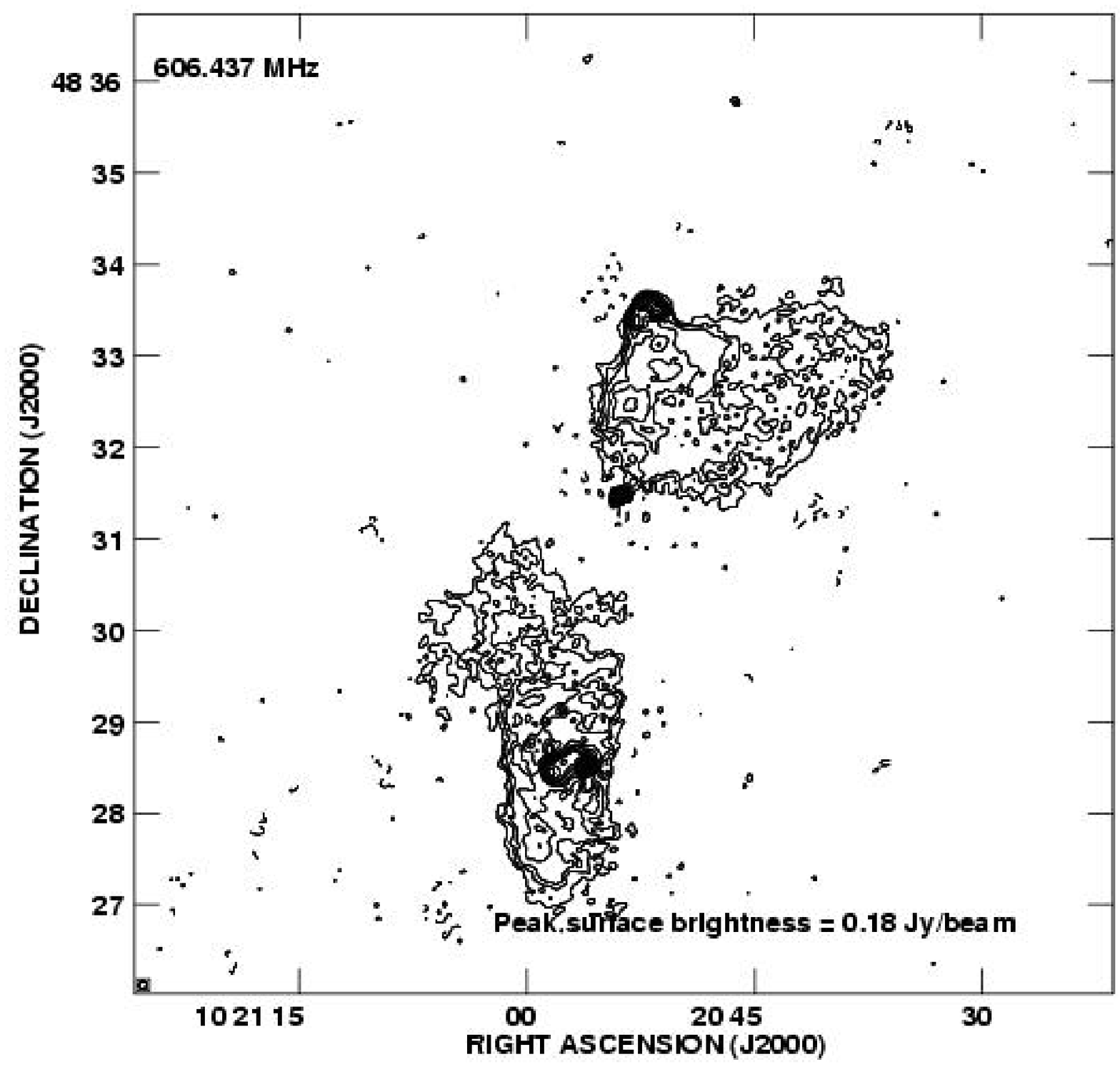} &
\includegraphics[width=7.5cm]{MAPS/4C48_29_240_PUB.PS} \\ [-1.2cm]
\includegraphics[width=7.5cm]{MAPS/4C48_29_MATCH_PUB.PS} &
\includegraphics[width=7.5cm]{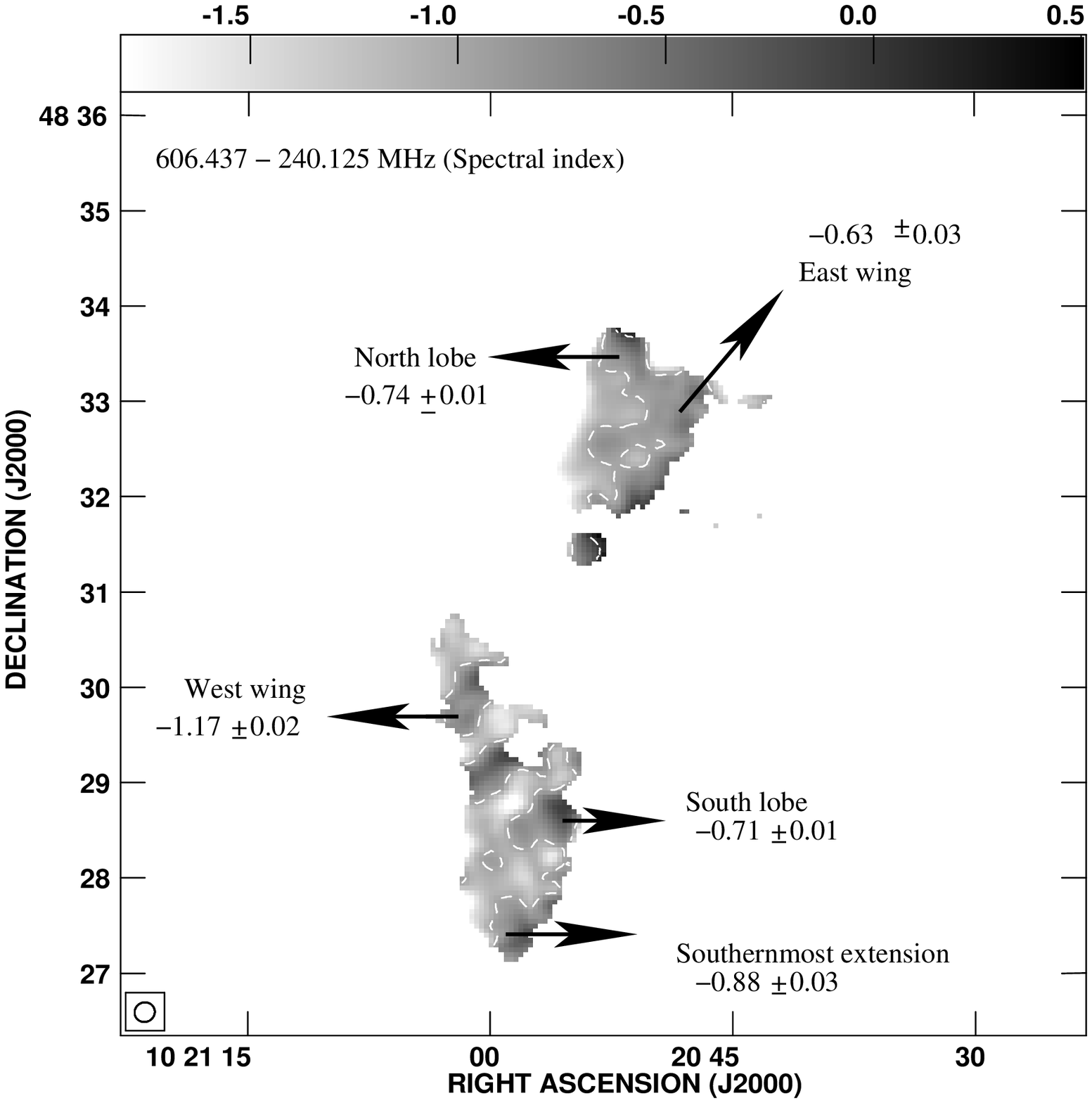} \\ [-0.8cm]
\end{tabular}
\end{center}
\caption{Upper: Full synthesis GMRT maps of 4C 48.29 at 610
(left panel) and 240~MHz (right panel).
The CLEAN beams for 610 and 240 MHz maps are
5$^{\prime\prime}$.1~$\times$~5$^{\prime\prime}$.0
at a P.A. of $-$65$^{\circ}$.3
and
12$^{\prime\prime}$.6~$\times$~12$^{\prime\prime}$.2
at a P.A. of $-$48$^{\circ}$.9,
respectively; and the contour levels in the two maps, respectively are
$-$0.8, 0.8, 1, 2, 3, 4, 5, 6, 8, 10, 40 mJy~beam$^{-1}$
and
$-$12, 12, 16, 24, 40, 60, 80, 100, 160, 200 mJy~beam$^{-1}$.
Lower left: The map of 4C 48.29 at 610~MHz
matched with the resolution of 240~MHz.
The contour levels are
$-$2, 2, 4, 6, 8, 10, 16, 20, 40, 60, 80, 100 mJy~beam$^{-1}$.
Lower right: The distribution of the spectral index,
between 240 and 610 MHz, for the source.
The spectral index contours are at $-$1,~0.
The error-bars in the full synthesis maps found at a source free location
are $\sim$2.6 and $\sim$0.2~mJy~beam$^{-1}$ at 240 and 610~MHz, respectively.
}
\label{full_syn_48_29}
\end{figure*}

\begin{figure*}
\begin{center}
\begin{tabular}{ll}
\includegraphics[width=7.5cm]{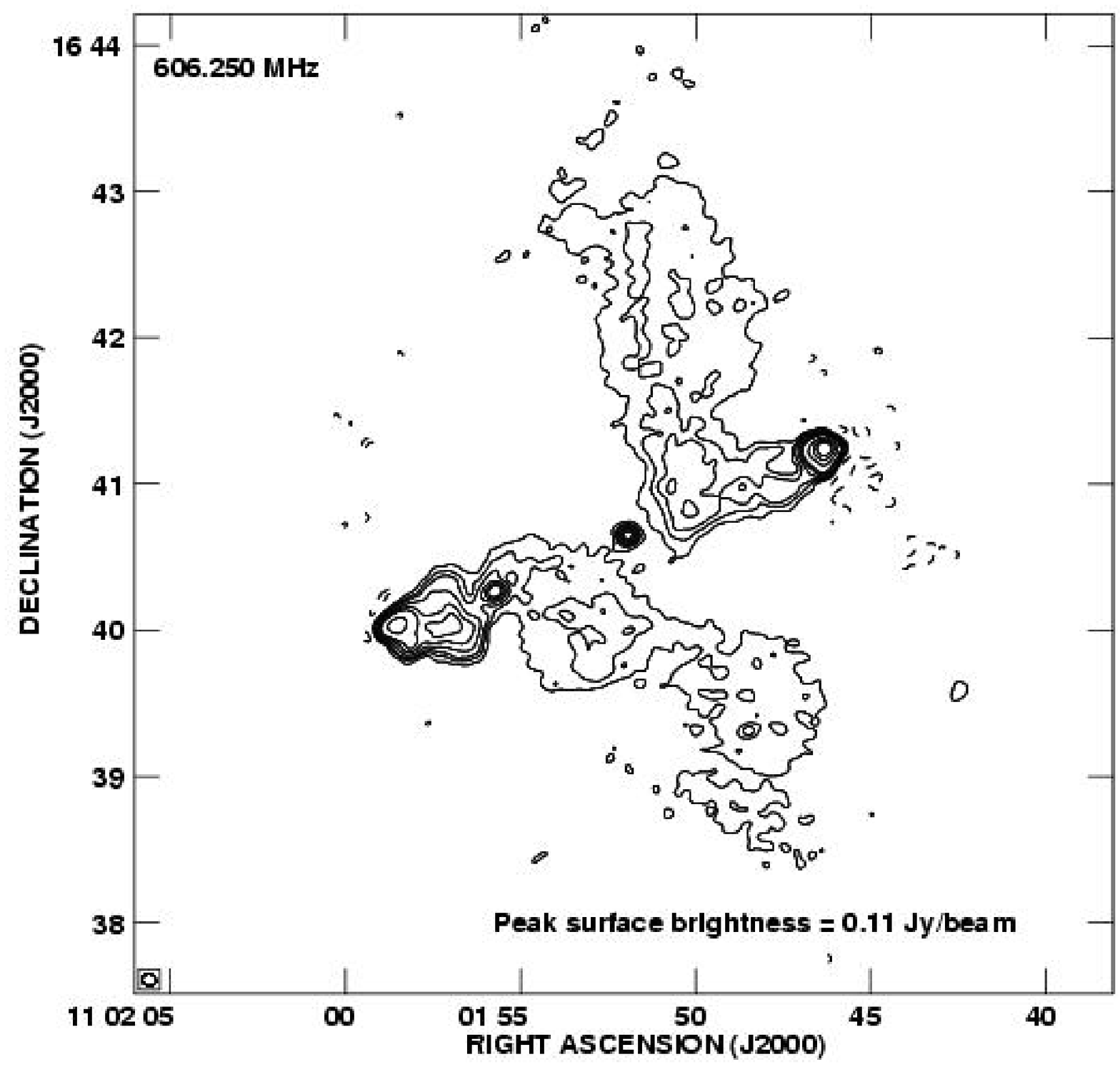} &
\includegraphics[width=7.5cm]{MAPS/B1059_169_240_PUB.PS} \\ [-1.2cm]
\includegraphics[width=7.5cm]{MAPS/B1059_169_MATCH_PUB.PS} &
\includegraphics[width=7.5cm]{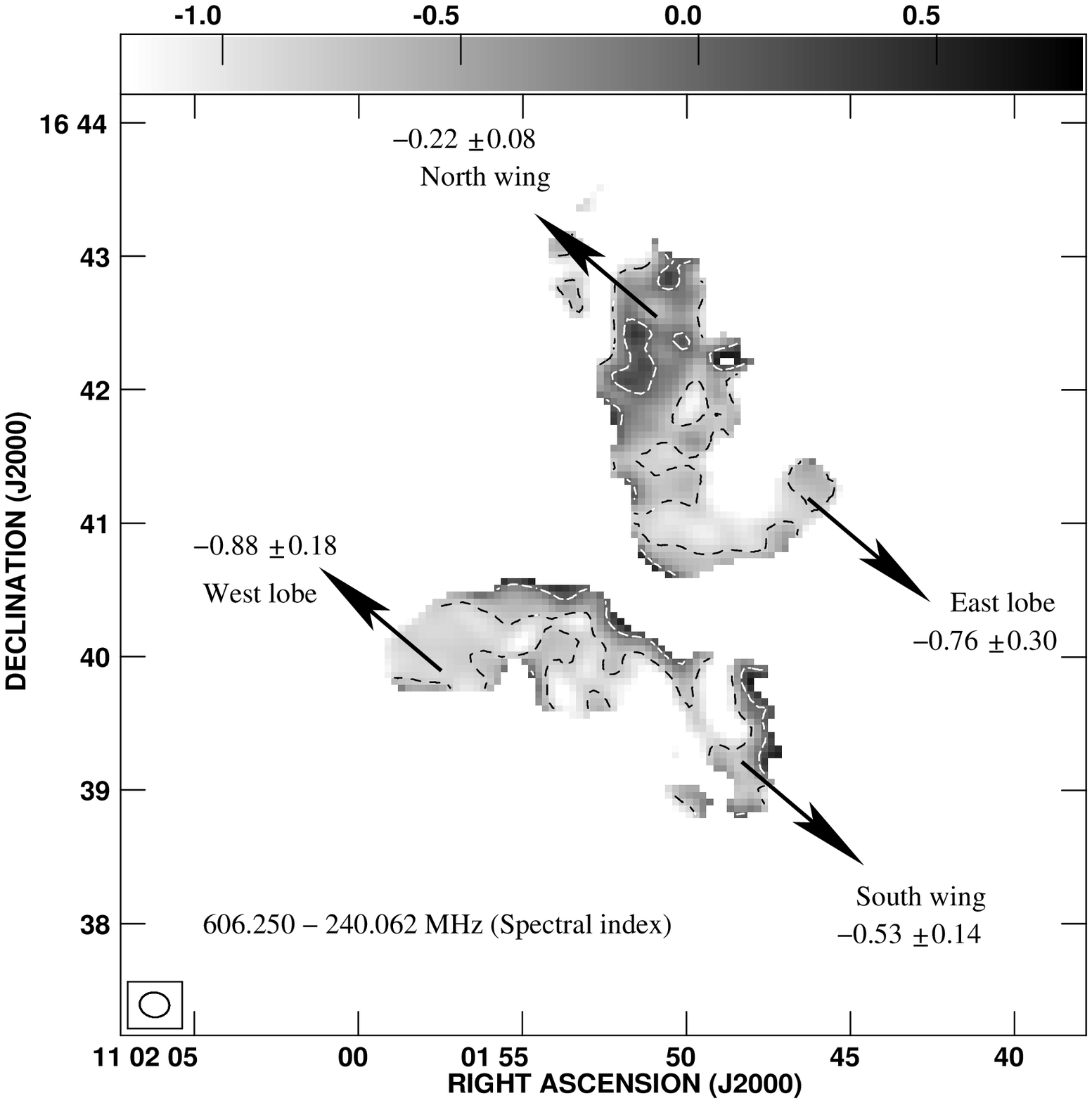} \\ [-0.8cm]
\end{tabular}
\end{center}
\caption{Upper: Full synthesis GMRT maps of B1059$+$169 at 610
(left panel) and 240~MHz (right panel).
The CLEAN beams for 610 and 240 MHz maps are
5$^{\prime\prime}$.7~$\times$~4$^{\prime\prime}$.9
at a P.A. of 77$^{\circ}$.0
and
13$^{\prime\prime}$.0~$\times$~11$^{\prime\prime}$.1
at a P.A. of 82$^{\circ}$.5,
respectively; and the contour levels in the two maps, respectively are
$-$1, 1, 2, 4, 6, 8, 16, 24, 32 mJy~beam$^{-1}$
and
$-$3, 3, 4, 6, 8, 10, 12, 20, 30, 40, 60, 100 mJy~beam$^{-1}$.
Lower left: The map of B1059$+$169 at 610~MHz
matched with the resolution of 240~MHz.
The contour levels are
$-$1, 1, 2, 3, 4, 6, 8, 12, 20, 40, 60 mJy~beam$^{-1}$.
Lower right: The distribution of the spectral index,
between 240 and 610 MHz, for the source.
The spectral index contours are at $-$0.8,~0.
The error-bars in the full synthesis maps found at a source free location
are $\sim$1.0 and $\sim$0.2~mJy~beam$^{-1}$ at 240 and 610~MHz, respectively.
}
\label{full_syn_b1059}
\end{figure*}

\begin{figure*}
\begin{center}
\begin{tabular}{ll}
\includegraphics[width=7.5cm]{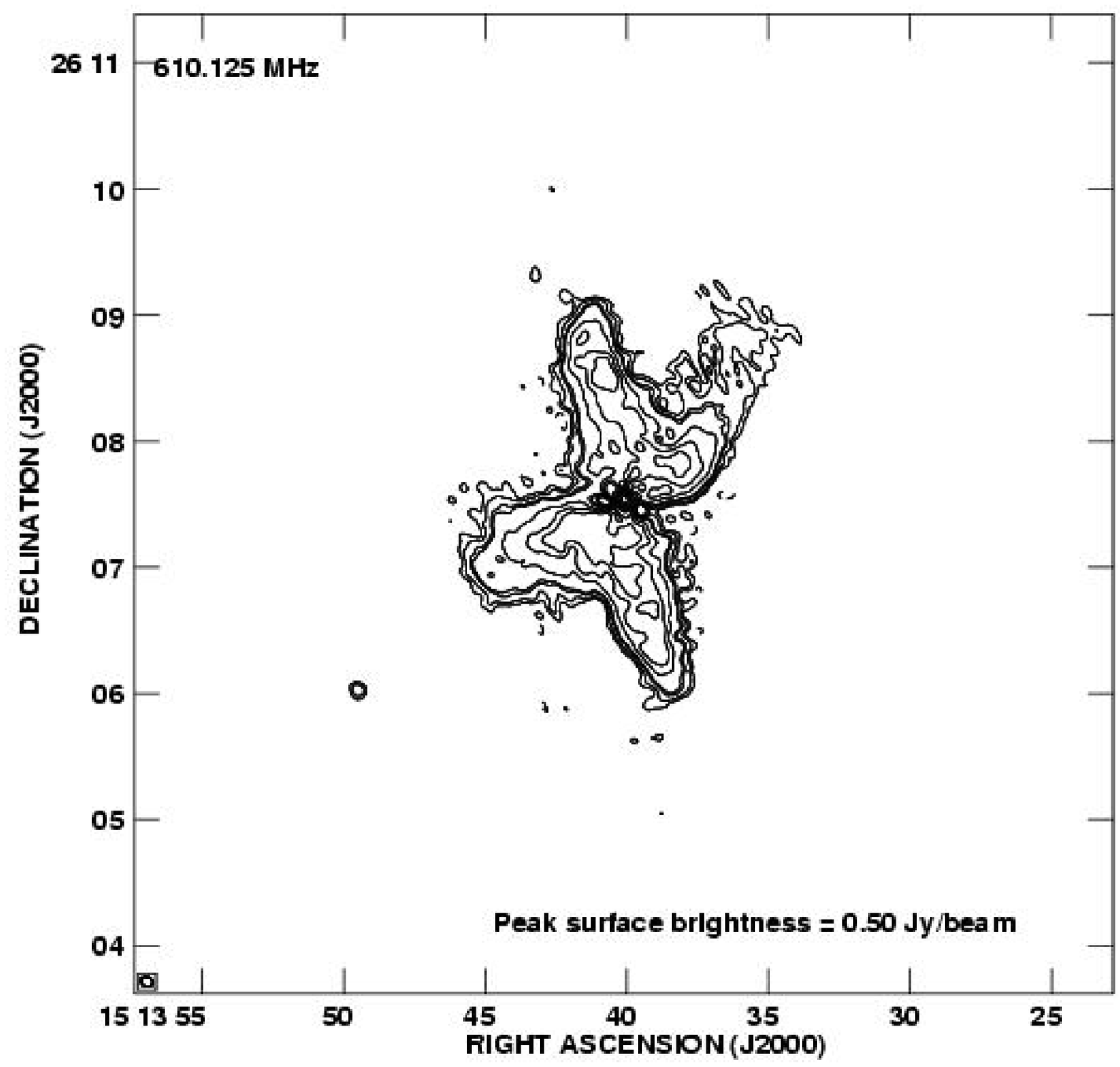} &
\includegraphics[width=7.5cm]{MAPS/3C315_240_PUB.PS} \\ [-1.2cm]
\includegraphics[width=7.5cm]{MAPS/3C315_MATCH_PUB.PS} &
\includegraphics[width=7.5cm]{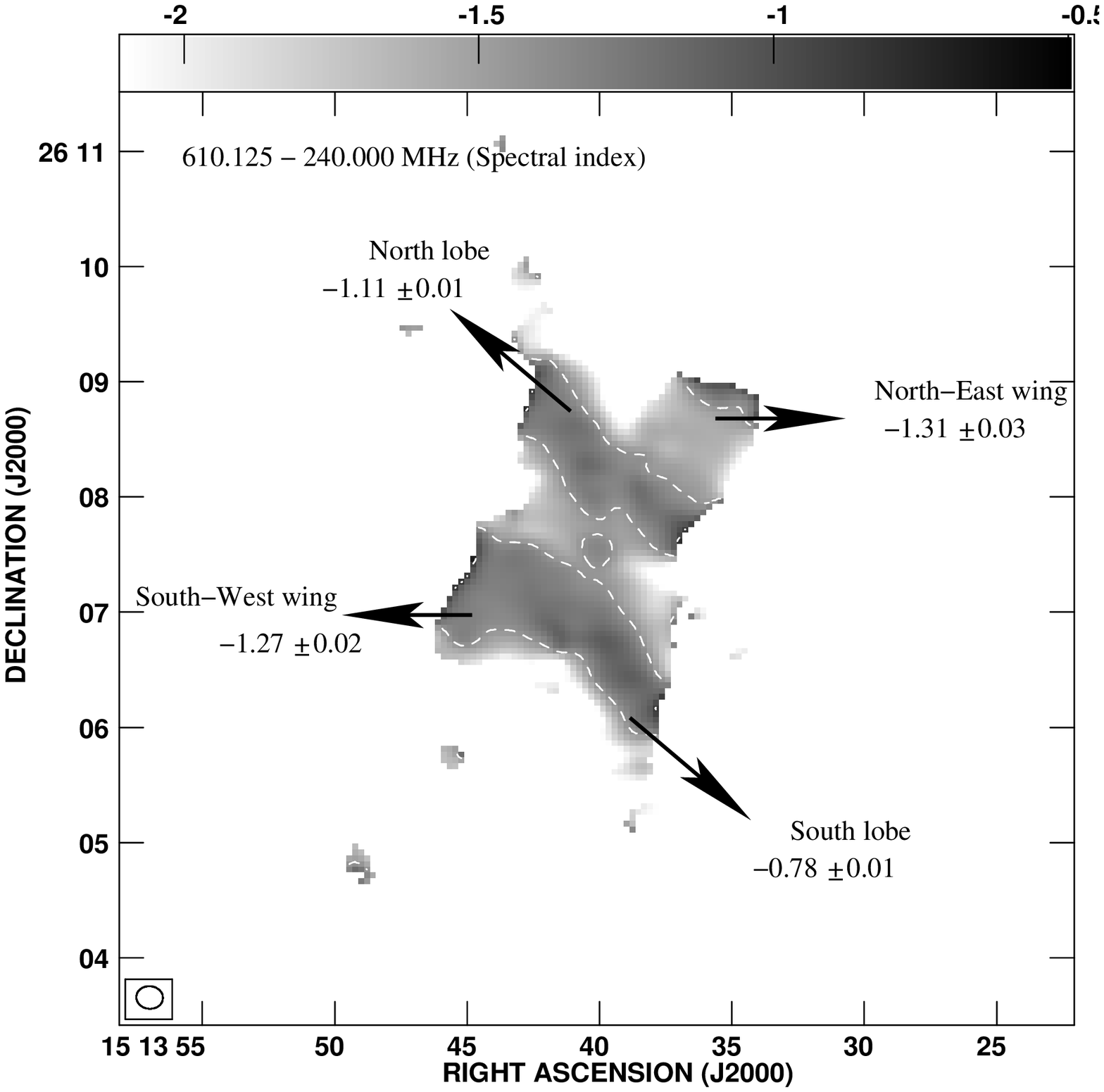} \\ [-0.8cm]
\end{tabular}
\end{center}
\caption{Upper: Full synthesis GMRT maps of 3C 315 at 610
(left panel) and 240~MHz (right panel).
The CLEAN beams for 610 and 240 MHz maps are
5$^{\prime\prime}$.7~$\times$~4$^{\prime\prime}$.7
at a P.A. of 62$^{\circ}$.0
and
13$^{\prime\prime}$.6~$\times$~11$^{\prime\prime}$.8
at a P.A. of 84$^{\circ}$.1,
respectively; and the contour levels in the two maps, respectively are
$-$2, 2, 4, 8, 20, 30, 40, 80, 100, 200, 400, 800 mJy~beam$^{-1}$
and
$-$160, 160, 240, 400, 500, 600, 800, 1000, 1200, 2000 mJy~beam$^{-1}$.
Lower left: The map of 3C 315 at 610~MHz
matched with the resolution of 240~MHz.
The contour levels are
$-$3, 3, 4, 8, 10, 20, 40, 100, 120, 160, 200, 240, 300 mJy~beam$^{-1}$.
Lower right: The distribution of the spectral index,
between 240 and 610 MHz, for the source.
The spectral index contours are at $-$1.7, 0.5.
The error-bars in the full synthesis maps found at a source free location
are $\sim$6.8 and $\sim$0.3~mJy~beam$^{-1}$ at 240 and 610~MHz, respectively.
}
\label{full_syn_315}
\end{figure*}

\subsubsection*{3C~192 (z = 0.060)}

3C~192 was classified, as a $X$-shaped source having prominent distorted
structure of rotational type \citep{Parmaetal}.

Fig.~\ref{full_syn_192} shows the radio images at 240 and 610~MHz.
The angular extents along the active axis and
the wings are $\sim$210\arcsec and $\sim$120\arcsec, respectively
at both frequencies.
Both the maps show similar radio extent along the active axis and along
the wings, consistent with earlier work \citep{Hogbom,Dennett1}, and
both the wings have symmetrical distortions.
The radio structure in the central parts of this source is confused,
but the detailed shape of the contours close to the position of the
host galaxy clearly indicates the presence of a weak unresolved
central component.  There is also a weak transverse feature across the
south-east wing.

The low frequency fitted spectra have
$-$0.82 $> \alpha >$ $-$1.15 for all regions across the source.
The source shows evidence for comparable spectra in the
active lobes than in the wings. The north-west and south-east wings have
spectral indices,
$-$0.88 $\pm$0.03 and $-$0.98 $\pm$0.02, respectively,
whereas the north-east and south-west
active lobes have $-$0.82 $\pm$0.01 and $-$1.15 $\pm$0.01, respectively.

\subsubsection*{B2~0828$+$32 (z = 0.053)}

B2~0828$+$32 is the most extended $X$-shaped source in the sample.
Although, the host galaxy neither have a double core nor a
companion, the luminosity profile shows faint possibility of a merger
event \citep{UlrichRonnback}.

Fig.~\ref{full_syn_b2} shows the radio images at 240 and 610~MHz.
The source shows large scale `S' type distortions, suggestive
of precession phenomenon of the central engine \citep{Parmaetal}.
The south-east--north-west axis consists of prominent hot~spots,
which is possibly the active axis; whereas the north-south axis is
along the wings, which is barely visible in our maps
and in our low resolution (30\arcsec $\times$ 30\arcsec) maps.
The faint detection of hot~spot is likely in our 610~MHz map
and not in our low resolution maps, which were
earlier detected in high resolution images at 1.4 and 5~GHz.
\citep{Parmaetal,Ferettietal}.
This is one of the first $X$-shaped source,
identified as two double radio
structures of different ages and oriented at widely different angles
\citep{UlrichRonnback}.

The high frequency spectral index (1.4--10.55~GHz) exhibits
gradual steepening from the active lobes ($\alpha \simeq$ $-$0.7)
towards the wings ($\alpha \simeq$ $-$1.1, southern wing
and $\alpha \simeq$ $-$1.1, northern wing) \citep{Rottmann}.
The low frequency fitted spectra have
$-$0.37 $\ge \alpha \ge$ $-$2.79
for all regions across the source.
Even though the radio maps and the spectral index map are
rather noisy, due mainly to the absence of the north wing and
the south wing in our 240~MHz map, there is a weak
evidence for steeper spectra in the
active lobes than in the wings. The spectral index
of the north wing is flatter than $\ge$$-$2.79 $\pm$0.30
and of the south wing is steeper than $\le$$-$0.37 $\pm$0.21.
Instead, the north-west and south-east active lobes have
spectral indices of $-$1.27 $\pm$0.01 and $-$1.21 $\pm$0.01,
respectively.

\subsubsection*{3C~223.1 (z = 0.108)}

The host galaxy was imaged as part of the {\it HST} snapshot survey
\citep{deKoff} and the galaxy has a strong central bulge and
a very pronounced dust disk.  The source is believed to be
isolated or in a poor group \citep{Sandage1972} and no X-ray emission
was detected from it or at its surroundings \citep{Burns1981}.

Fig.~\ref{full_syn_223_1}
shows complex radio source with an $X$-shaped morphology at both
240 and 610~MHz. The angular extent is $\sim$105\arcsec along
the active lobes (those with hot-spots) and
$\sim$150\arcsec along the wings.
The nuclear source of 3C~223.1 is invisible at both these frequencies
and also in the radio maps of \citet{Dennett},
but is detected and is unresolved at 8.4~GHz \citep{Blacketal}.
The weak jet detected mid-way between core and north lobe
at 8.4~GHz \citep{Blacketal} is not seen in our maps,
because of coarser resolution.
Our maps also suggest of a sharp boundary at the
farthest end of the north lobe and a likely ring-like feature in the
south lobe, which is consistent with earlier results.

The low frequency fitted spectra have
$-$0.37 $> \alpha >$ $-$1.08 for all regions across the source, and
is the first $X$-shaped source
showing evidence for steeper spectra in the
active lobes than in the wings \citep{LalRao2005}.
Here, we do look into the errors that would be introduced due to
possible negative depression. The maximum depression close
to the source at 240 and 610 MHz are
$-$7.6 and $-$2.7~mJy~beam$^{-1}$ respectively.
This worst case would introduce a maximum error of 0.14 in spectral
indices for the wings and 0.01 for the active lobes.
The east and west wings have spectral indices,
$-$0.37 $\pm$0.14 and $-$0.62 $\pm$0.14, respectively,
whereas the north and south
active lobes have $-$1.08 $\pm$0.01 and $-$1.08 $\pm$0.01, respectively, and
are consistent with spectral results between 1.4 and 32~GHz, {\it i.e.}
$-$0.70 $\pm$0.03 and $-$0.66 $\pm$0.03 for the east and west wings, and
$-$0.75 $\pm$0.02 and $-$0.77 $\pm$0.02 for the north and south active lobes,
respectively \citep{Dennett}.
Since, the observed differences in spectral index at low frequencies
are much more than the uncertainties,
we believe that the observed spectral index features are real.
Similar result, `spectral reversal' was also found independently by
\citet{Rottmann}.

\subsubsection*{4C~48.29 (z = 0.053)}

This source is the nearest to us and is at the centre of Abell~990 cluster.
The parent galaxy belongs to a double system with no visible companion
on the Palomar Sky Survey (PSS) prints \citep{Parmaetal}, but a companion
is detected, $\sim$25\arcsec away on the south-east
on the 2~Metre All Sky Survey (2MASS), which is probably a starburst galaxy.

Fig.~\ref{full_syn_48_29} shows the radio images at 240 and 610~MHz and
they show that the brightest components of the hot~spots
are well aligned with the core, which is consistent with the
results of \citet{vanBreugelJagers}.  The source has a peculiar $X$-shaped
morphology, with an additional,
wing like feature at the south of the south active lobe.
Furthermore, 240~MHz map shows a low surface brightness feature $\sim$2\arcmin
to the east and a point source $\sim$4\arcmin to the south-east of the core.

The low frequency fitted spectra have
$-$0.63 $> \alpha >$ $-$1.17 for all regions across the source.
The east and west wings have spectral indices,
$-$0.63 $\pm$0.03 and $-$1.17 $\pm$0.02, respectively,
whereas the north and south
active lobes have $-$0.74 $\pm$0.01 and $-$0.71 $\pm$0.01, respectively.
Furthermore, the southernmost extension has a
spectral index of $-$0.88 $\pm$0.03.
Briefly, the east wing definitely shows evidence for flatter spectra than
either of the two active lobes;
otherwise rest of the regions have comparable spectral indices.

\subsubsection*{B1059$+$169 (z = 0.068)}

B1059$+$169 is the another source seen in cluster environment (Abell~1145),
which is the dominant radio galaxy
and is $\sim$5.5\arcmin away from the cluster centre.
A companion is detected on the 2MASS,
coincident with the cluster centre.

The morphology at low frequency (Fig.~\ref{full_syn_b1059}) is similar
to the 1.4~GHz map of \citet{OwenLedlow}.
The extent of the east-west active axis is 200\arcmin ~in
both the maps, whereas the extent of the north-south
axis in 240~MHz map is 310\arcmin,
which is slightly more than the extent of the wings in the 610~MHz map.
This could be possibly due to spectral ageing,
which is a common phenomenon for head-tail and wide-angle-tail radio sources,
seen in clusters, exhibiting low-surface-brightness features.
In the 610~MHz map, we also detect a knot inbetween the west lobe and the core.

The low frequency fitted spectra have
$-$0.22 $> \alpha >$ $-$0.88 for all regions across the source.
Although B1059$+$169 is found in the cluster environment,
surprisingly, the spectral results of it are similar to that of 4C~12.03
and 3C~223.1, {\it i.e.} the wings have relatively
flatter spectral index as compared to the active lobes.
The north and south wings have spectral indices,
$-$0.22 $\pm$0.08 and $-$0.53 $\pm$0.14, respectively,
whereas the east and west
active lobes have $-$0.76 $\pm$0.30 and $-$0.88 $\pm$0.18, respectively.
Furthermore, the knot inbetween the west lobe and the core has a spectral
index of $-$0.82 $\pm$0.22, similar to the east and west active lobes.

\begin{figure*}
\begin{center}
\begin{tabular}{ll}
\includegraphics[width=7.5cm]{MAPS/3C403_610_PUB.PS} &
\includegraphics[width=7.5cm]{MAPS/3C403_240_PUB.PS} \\ [-1.2cm]
\includegraphics[width=7.5cm]{MAPS/3C403_MATCH_PUB.PS} &
\includegraphics[width=7.5cm]{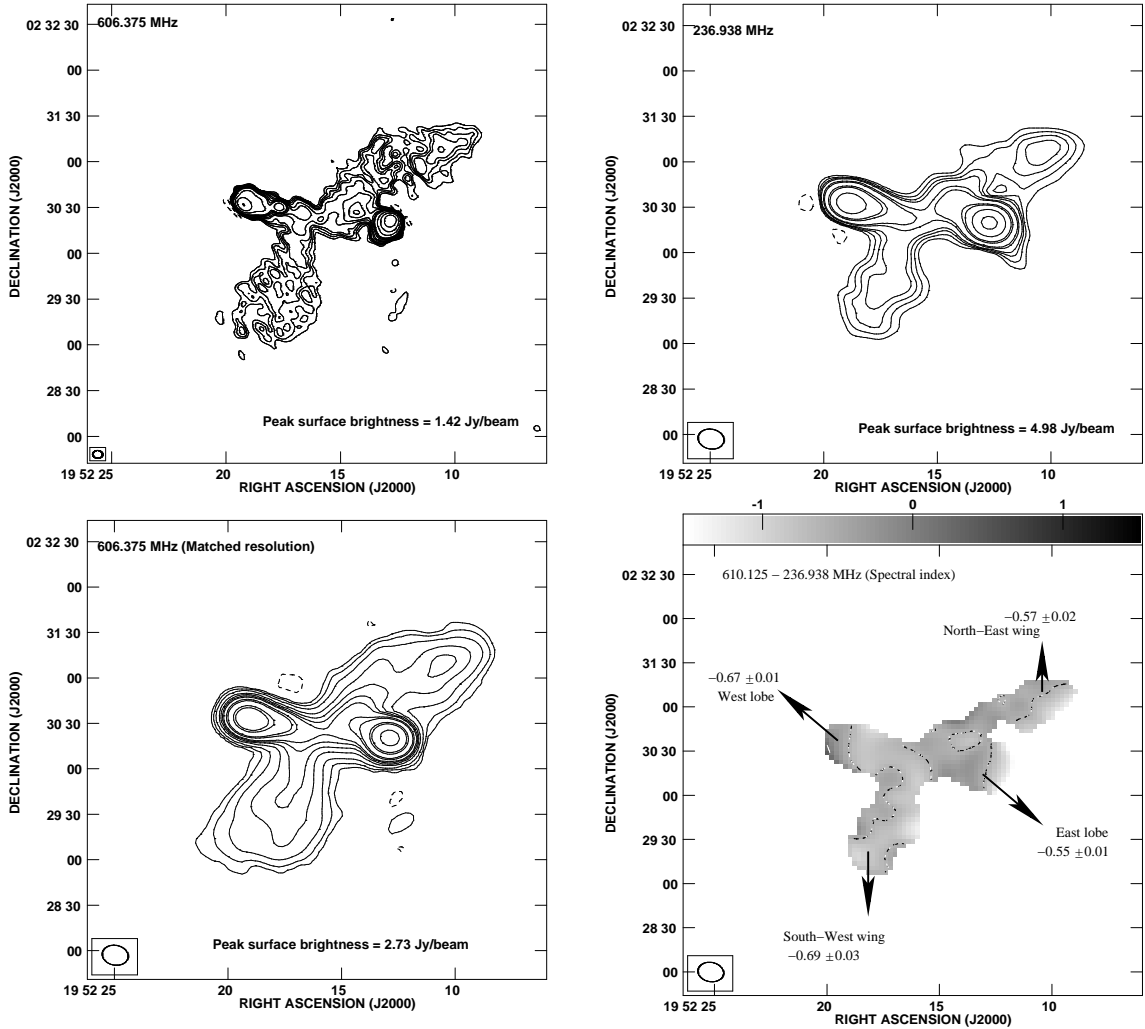} \\ [-0.8cm]
\end{tabular}
\end{center}
\caption{Upper: Full synthesis GMRT maps of 3C 403 at 610
(left panel) and 240~MHz (right panel).
The CLEAN beams for 610 and 240 MHz maps are
6$^{\prime\prime}$.4~$\times$~5$^{\prime\prime}$.1
at a P.A. of 85$^{\circ}$.3
and
17$^{\prime\prime}$.3~$\times$~13$^{\prime\prime}$.1
at a P.A. of 79$^{\circ}$.0,
respectively; and the contour levels in the two maps, respectively are
$-$8, 8, 12, 16, 20, 30, 40, 60, 80, 100, 200, 400, 800 mJy~beam$^{-1}$
and
$-$80, 80, 120, 160, 200, 320, 400 800 1000 mJy~beam$^{-1}$.
Lower left: The map of 3C 403 at 610~MHz
matched with the resolution of 240~MHz.
The contour levels are
$-$10, 10, 20, 40, 80, 120, 160, 200, 320, 400, 800, 1000 mJy~beam$^{-1}$.
Lower right: The distribution of the spectral index,
between 240 and 610 MHz, for the source.
The spectral index contours are at $-$0.6, 0.2.
The error-bars in the full synthesis maps found at a source free location
are $\sim$4.8 and $\sim$0.5~mJy~beam$^{-1}$ at 240 and 610~MHz, respectively.
}
\label{full_syn_403}
\end{figure*}

\begin{figure*}
\begin{center}
\begin{tabular}{ll}
\includegraphics[width=7.5cm]{MAPS/3C433_610_PUB.PS} &
\includegraphics[width=7.5cm]{MAPS/3C433_240_PUB.PS} \\ [-1.2cm]
\includegraphics[width=7.5cm]{MAPS/3C433_MATCH_PUB.PS} &
\includegraphics[width=7.5cm]{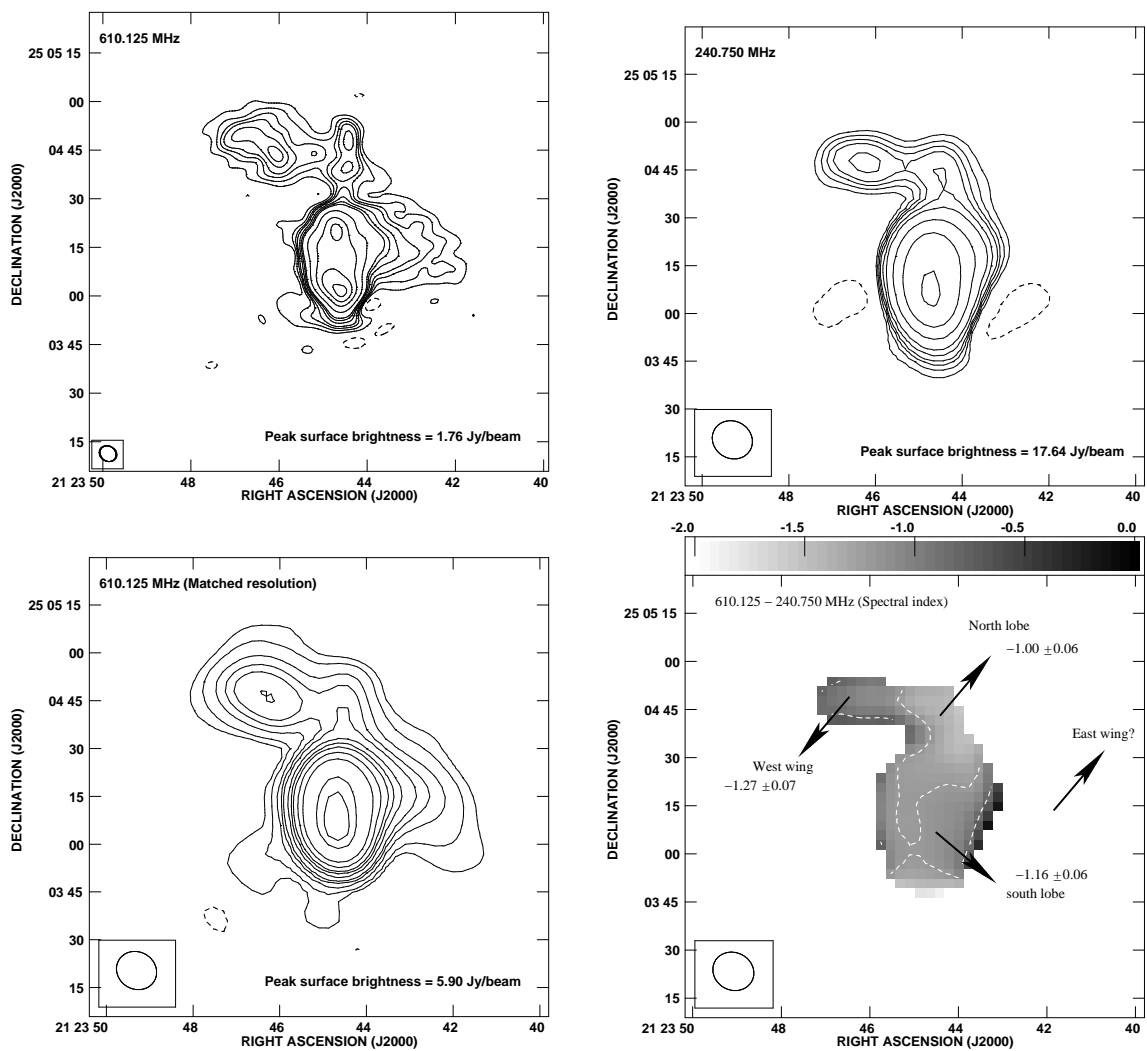} \\ [-0.8cm]
\end{tabular}
\end{center}
\caption{Upper: Full synthesis GMRT maps of 3C 433 at 610
(left panel) and 240~MHz (right panel).
The CLEAN beams for 610 and 240 MHz maps are
5$^{\prime\prime}$.4~$\times$~4$^{\prime\prime}$.6
at a P.A. of 57$^{\circ}$.9
and
12$^{\prime\prime}$.9~$\times$~11$^{\prime\prime}$.6
at a P.A. of 58$^{\circ}$.7,
respectively; and the contour levels in the two maps, respectively are
$-$20, 20, 40, 60, 80, 120, 160, 200, 400, 800, 1200 mJy~beam$^{-1}$
and
$-$400, 400, 600, 800, 1000, 1400, 2000, 4000, 8000 mJy~beam$^{-1}$.
Lower left: The map of 3C 433 at 610~MHz
matched with the resolution of 240~MHz.
The contour levels are
$-$50, 50, 100, 200, 300, 400, 600, 800, 1000, 1400, 2000, 4000 mJy~beam$^{-1}$.
Lower right: The distribution of the spectral index,
between 240 and 610 MHz, for the source.
The spectral index contours are at $-$1.2, $-$0.8, 0.
The error-bars in the full synthesis maps found at a source free location
are $\sim$6.9 and $\sim$0.8~mJy~beam$^{-1}$ at 240 and 610~MHz, respectively.
}
\label{full_syn_433}
\end{figure*}

\subsubsection*{3C~315 (z = 0.108)}

Similar to NGC~326 and 4C~48.29, here also, the host galaxy belongs to
a double system with no visible companion on the PSS prints
\citep{Parmaetal}, but $\sim$7\arcsec away on the south,
a companion is detected on the 2MASS,
and is identified to be a irregular galaxy.
Their optical profiles show signs of interaction and
seems to be located in a poor cluster \citep{Zirbel1997}.
{\it HST} image of the northern core, containing the radio source
shows an elliptical host \citep{deKoff}.

Fig.~\ref{full_syn_315} shows the radio images at 240 and 610~MHz with
both the axes to be roughly of similar angular sizes and surface
brightnesses, making it a very unusual source \citep{AlexanderLeahy1987}.
But, as suggested by \citet{Hogbom}, the jet along the north-south axis
being of slightly higher surface brightness and could define a recent
axis of activity. The core is clearly detected in both the radio maps
and is associated with the brightest member of a pair of
elliptical galaxy \citep{LeahyWilliams}.
Furthermore, the source does not display characteristic elongated structure,
{\it i.e.} the usual hot~spots at or near the extreme edges.

The spectral index map of 3C~315 using the radio maps at 1.65 and 2.7~GHz
at a resolution of 9.1\arcsec $\times$ 5.5\arcsec
shows relatively flat spectral index regions close to the core and
steep spectral index regions being located at the tips of the active lobes.
The southern and northern active lobes have high frequency (1.65--2.7~GHz)
spectral indices, $-$1.46 and $-$1.26, respectively \citep{Rottmann}.
The low frequency fitted spectra have
$-$0.78 $> \alpha >$ $-$1.31 for all regions across the source, and
shows evidence for flatter spectra in the
active lobes than in the wings.
The north-east and south-west wings have
spectral indices, $-$1.31 $\pm$0.03 and $-$1.27 $\pm$0.02, respectively,
whereas the north and south
active lobes have $-$1.11 $\pm$0.01 and $-$0.78 $\pm$0.01, respectively.
The spectral index map (Fig. \ref{full_syn_315}, lower right panel)
shows peculiar spectral
behaviour and are consistent with the findings of
\citet{AlexanderLeahy1987} and \citet{Rottmann}, {\it i.e.}
the spectrum is steep at the wings,
it becomes flatter in regions close to the core,
and again steepens towards the active lobe and ultimately
becomes steepest at the tip of the active lobes.

\subsubsection*{3C~403 (z = 0.059)}

The host of 3C~403 is a E0, narrow-line radio galaxy (NED classification)
and the continuum colors are typical of an early type galaxy.
The galaxy appears to be a smooth elliptical using {\it HST}
with an apparent separation of the source into a central elliptical region and
a low-surface-brightness halo, which is probably due to intervening dust
\citep{Martel1999}.
The host galaxy does not have any bright companion and
is located in a very low-density local galaxy environment \citep{Heckman1994}.

Fig.~\ref{full_syn_403} shows the radio images at 240 and 610~MHz, and
the east and west hot~spots are clearly detected.
The slightly larger angular extent of the north-south axis along the wings
in 240~MHz map than in 610~MHz and high frequency maps \citep{Dennett1}
could be due to spectral ageing.

The low frequency fitted spectra have
$-$0.55 $> \alpha >$ $-$0.69 for all regions across the source.
The south-west and north-east wings have
spectral indices, $-$0.69 $\pm$0.03 and $-$0.57 $\pm$0.02, respectively,
whereas the east and west active lobes have
$-$0.55 $\pm$0.01 and $-$0.67 $\pm$0.01, respectively.
These results are inconsistent with
results at high frequency \citep{Dennett,Rottmann},
{\it i.e.} the spectral indices between 1.4 and 32~GHz are
$-$0.80 $\pm$0.03 and $-$0.77 $\pm$0.28
for the south-west and north-east wings, respectively,
(note that \citealt{Dennett} label the south-west and north-east wings
as the south-east and north-west wings, respectively), and
$-$0.78 $\pm$0.02 and $-$0.77 $\pm$0.02 for the east and west active lobes,
respectively.
Instead, the source shows evidence for comparable spectra
in the active lobes and the wings.

\subsubsection*{3C~433 (z = 0.102)}

The host has a disturbed morphology (PSS prints)
with no known nearby companion
and lies in a cluster \citep{McCarthy1995}.
{\it HST} image the source
shows galaxy full of filaments of dust, a faint patch of
emission north-west of the galaxy coinciding with a spot of
radio emission and possible regions of star formation \citep{deKoff}.

Fig.~\ref{full_syn_433} shows the radio images at 240 and 610~MHz.
Since, our map at 240~MHz is a low resolution map, therefore
several features seen in \citet{vanBreugeletal} and \citet{Blacketal}
are smoothed, and it is unclear if the source is indeed $X$-shaped source.
Instead, the 610~MHz map shows many of the features to be clearly
resolved similar to  the VLA map at 6~cm \citep{vanBreugeletal,Blacketal}.
We clearly detect weak, collimated emission to the north from the lobe
forming, A, B and C \citep[for the positions of A to I,
see Fig. 1(a) of][]{vanBreugeletal}.
The eastern feature, D is also clearly detected.
As suggested by \citet{vanBreugeletal},
the south lobe seem to flare out close to the nucleus with a
relatively large opening angle forming the eastern jet along I.
We define regions centered at B and H to be the north and south active lobes,
respectively, and we believe that the two lobes do not appear to be
relaxed systems. The low-surface-brightness regions centered at D and I,
respectively, are assumed to be the east and west wings
for our further analysis.

We do not detect the east wing in our low frequency map
and the fitted spectra have
$-$1.00 $> \alpha \ge$ $-$3.42 for all regions across the source.
The source shows marginal evidence for comparable spectra in the
active lobes and the wings. The west wing has
spectral index, $-$1.27 $\pm$0.03 and
the east wing is flatter than $\ge$$-$3.42 $\pm$0.21,
whereas the north and south active lobes have
$-$1.00 $\pm$0.03 and $-$1.16 $\pm$0.01, respectively.
Here again, we quantify the errors introduced due to
possible negative depression,
being $-$94.2 and $-$10.1~mJy~beam$^{-1}$ at 240 and 610 MHz respectively,
close to the source;
thereby introducing a maximum error of 0.07 in spectral
indices for the wings and 0.06 for the active lobes, which are smaller
than the quoted statistical errors.

\begin{figure}
\begin{center}
\begin{tabular}{l}
\includegraphics[width=8.1cm]{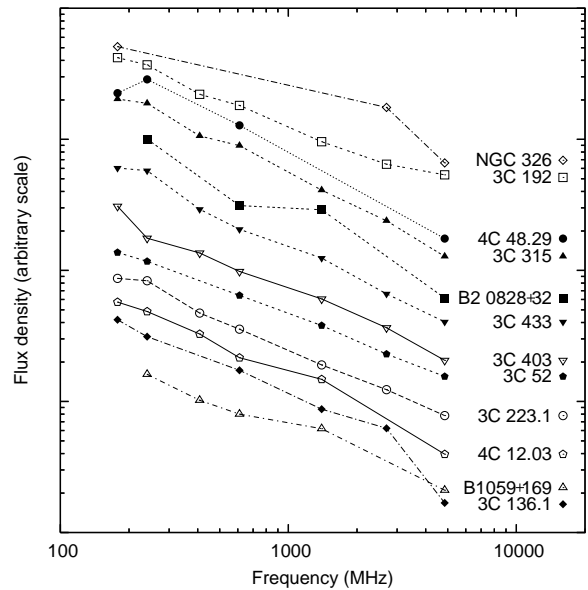}
\end{tabular}
\end{center}
\caption{Integrated flux density (spectra) for the sample of
$X$-shaped sources.  Various measurements along with the error-bars
(not plotted) are explained in Table~\ref{flux_density}.
The spectra are shifted with respect to one another for clarity.
}
\label{flux_reg}
\end{figure}

\section{Discussion} 
\label{discussion}

The low frequency GMRT observations combined with observations at other
wavelengths raise a number of questions regarding
the nature and the formation scenario of the $X$-shaped sources.
Some of the earlier scenarios that were based on limited information
need to be investigated. In this section we first, 
discuss the overall source morphological
and spectral properties (Section~\ref{overall}).
We then weigh all the {\it pros and cons} of existing models
(Section~\ref{form_mod}), and statistical implications of these
results on our understanding of the current formation models
for the known sample of $X$-shaped sources (Section~\ref{implications}).
Finally, in Section~\ref{single}, we address the question,
do $X$-shaped sources constitute a single class in the light of
`merger of two AGNs' model.

\begin{table}
\caption{Flux densities of all the distinct regions at 240 and 610 MHz.
The error-bars for these sources are quoted in the respective figure captions.
Label denotes: $^\dag$The upper limits on
the flux densities quoted are five times the rms noise levels.}
\centering
\begin{tabular}{l|lrr}
\hline \\
       &  & \multicolumn{2}{c}{Flux density} \\
       &  & \multicolumn{1}{c}{240 MHz} & \multicolumn{1}{c}{610 MHz} \\
       &  & \multicolumn{2}{c}{(mJy)} \\
\hline \\
4C~12.03    &North lobe      & 2214.9    & 915.0  \\
            &South lobe      & 1376.3    & 618.9  \\
            &East  wing      &  117.5    &  67.8  \\
            &West  wing      &   91.5    &  47.6  \\
3C~52       &North lobe      & 5312.0    &3035.9  \\
            &South lobe      & 2395.0    &1314.4  \\
            &East  wing      &  384.0    & 121.2  \\
            &West  wing      &  272.6    & 101.8  \\
3C~136.1    &East  lobe      &  904.0    & 511.3  \\
            &West  lobe      & 2149.0    &1214.6  \\
            &North wing      &  163.5    &  80.2  \\
            &South wing      &  159.5    &  82.4  \\
3C~192      &North-East lobe &  383.2    & 167.9  \\
            &South-West lobe &  478.7    & 192.0  \\
            &South-East wing & 2351.3    &1091.5  \\
            &North-West wing & 2405.8    & 824.8  \\
B2~0828$+$32&South-East lobe & 1288.8    & 416.8  \\
            &North-West lobe & 1213.4    & 372.0  \\
            &North wing      & $<$13.5$^\dag$  &$<$1.0$^\dag$  \\
            &South wing      & $<$13.5$^\dag$   &  19.1  \\
3C~223.1    &North lobe      & 2946.1    &1074.9  \\
            &South lobe      & 3268.6    &1196.7  \\
            &East  wing      &  123.8    &  87.8  \\
            &West  wing      &  175.5    &  98.3  \\
4C~48.29    &North lobe      &  462.2    & 232.3  \\
            &South lobe      &  689.7    & 356.8  \\
            &East  wing      &   51.9    &  28.9  \\
            &West  wing      &   98.9    &  33.2  \\
            &Southernmost    &  251.5    & 110.4  \\
B1059$+$169 &East  lobe      &  210.3    &  92.4  \\
            &West  lobe      &  226.7    & 111.6  \\
            &North wing      &   25.9    &  21.0  \\
            &South wing      &   23.5    &  14.4  \\
3C~315      &North lobe      & 1140.6    & 403.6  \\
            &South lobe      &  938.2    & 453.7  \\
            &North-East wing & 1018.3    & 300.6  \\
            &South-West wing &  977.5    & 291.7  \\
3C~403      &East  lobe      & 5856.9    &3133.0  \\
            &West  lobe      & 3740.9    &2241.7  \\
            &North wing      &  512.1    & 300.1  \\
            &South wing      &  549.5    & 287.9  \\
3C~433      &North lobe      & 1685.3    & 663.5  \\
            &South lobe      &21025.0    &7066.3  \\
            &East  wing      & $<$34.0$^\dag$   & 831.1  \\
            &West  wing      & 1530.8    & 467.0  \\
\\
\hline
\end{tabular}
\label{x_age}
\end{table}

\subsection{Overall source properties}
\label{overall}

Nearly a dozen sources have been mapped in detail, with a resolution
approaching to $\sim$1 kpc in most cases. Here, a few qualitative points about
the overall source morphological and spectral properties,
based on images presented are made.

\subsubsection{Source morphological properties}

The salient features of morphology of all $X$-shaped sources are as follows:

\begin{enumerate}
\item Almost all sources show similar angular sizes for the two axes.
Marginal differences between two axes in 3C~192 and 3C~403,
can be attributed to the projection effects.
\item Source 3C~315 have all four lobes of almost similar surface brightness,
thereby making difficult to identify the active lobes and the wings. 
\item Although, 3C~433 is classified as $X$-shaped source,
it does not have `typical' `$X$' shape morphology. It also is the source with
smallest angular size in the sample.
\item The radio core is detected for only one of the source,
B1059$+$169 in its 610 MHz map.
Whereas the radio maps of sources, 4C12.03, 4C~48.29 and 3C~315 at 610 MHz
show marginal detection of the radio cores.
\item Hot spots are almost always detected in the high-surface-brightness
active lobes in both, 610 and 240 MHz radio maps.
\item We detect the high-surface-brightness jets in almost, 
infact all the $X$-shaped sources and
this nearly 100\%, high rate of jet detection is similar to
\citet{Leahyetal97} and \citet{Hardcastleetal}
for classical radio sources.
\end{enumerate}

\subsubsection{Global spectral morphological properties}

The integrated radio flux densities at several locations across
each of these sources, allow us to group them into following three
categories: (A) Sources in which the wings are of flatter spectral indices
than the active lobes, {\it i.e.} 4C~12.03, 3C~223.1
and B1059$+$169. These sources have spectral indices,
$-$0.22 $> \alpha_{\rm wings} >$ $-$0.70 and
$-$0.76 $> \alpha_{\rm active~lobes} >$ $-$1.08
for all regions across the source. In addition,
in the former two sources,
the wings (or low-surface-brightness jets) have flatter spectral
indices with respect to the high-surface-brightness jets
at high radio frequencies \citep{Dennett,Rottmann},
consistent with our low frequency results.
(B) Sources in which the wings and the active lobes have
comparable spectral indices, {\it i.e.} 3C~192, B2~0828$+$32,
4C~48.29 and 3C~403.
Here, these classes of sources have spectral indices,
$-$0.37 $\ge$ $\alpha_{\rm wings}$ $\ge$ $-$2.79 and
$-$0.55 $> \alpha_{\rm active~lobes} >$ $-$1.27
for all regions across the source.
It therefore seems that
although the two regions, the wings and the active lobes,
have comparable spectral indices, they are relatively flat as compared
to typical steep spectrum features.
It is important to note, since we have limits for spectral indices
of the north and south wings, we have included B2~0828$+$32
in this category, although the south wing seems to be flatter than
the active lobes.
(C) Sources in which the wings are of steeper spectral indices
as compared to the active lobes, {\it i.e.} NGC~326, 3C~52,
3C~136.1, 3C~315 and 3C~433.
This category of sources have spectral indices,
$-$0.71 $> \alpha_{\rm wings}$ $\ge$ $-$3.42 and
$-$0.56 $> \alpha_{\rm active~lobes} >$ $-$1.16
for all regions across the source. This latter class of sources
are similar to typical radio galaxies, where low-surface-brightness
features have steeper spectral indices as compared to the
high-surface-brightness features, which have relatively flat
spectral indices.  The result of the former two classes of
sources are unusual, and we believe that it is not due to
possible artefacts, {\it e.g.}
different UV coverages,
images containing negative depression around the source,
image misalignment at 240 and 610 MHz, etc.
Also, these results do not support the known formation scenarios,
discussed below, in which the wings are interpreted as relics of past
radio jets and the active lobes as the newer ones.

Furthermore, the results from this class of sources is consistent with
earlier results that
there is no trend of spectral index with jet side at any brightness level,
which has been a subject of considerable debate
\citep{Dennett1,Dennett,Blacketal}.

\subsection{Formation models}
\label{form_mod}

The most intriguing fact about $X$-shaped radio galaxies is the
apparently small number of sources of $X$-shape.
Two possible explanations for the small number are,
(i) they are very exotic objects that form only rarely and
under extraordinary circumstances, and
(ii) they are normal radio galaxies that are currently in a short-lived
and/or rare phase of their evolution \citep{Rottmann}.
In order to explain these, it is necessary to
understand the formation process responsible for the
$X$-shaped phenomenon. We briefly introduce the key formation mechanisms
and discuss them in the light of existing observational results.

\subsubsection{Backflow}

\citet{LeahyWilliams} have argued in favour of backflow being
responsible for the formation of the wings of $X$-shaped sources.
Backflow is formed by jet material that is released by the
hot~spots and is then streaming back towards the host galaxy.
In the model, the backflow material remains collimated until
it meets the backflow from the opposite hot~spot and expands
laterally into a fat disk oriented perpendicular to the radio lobe axis.
For physical conditions prevailing in the lobes of radio galaxies,
with typical advance speed of the hotspots of a few per cent of $c$,
high apparent backflow speeds have been inferred
with roughly constant ratio of the speed of advance and
the speed of backflow \citep{Scheuer95}.
Assuming backflow velocities close to magnetosonic sound speed,
though real flows occur at lower speeds,
and typical secondary lobe lengths of a few hundred kiloparsec,
one requires a duration of the order of a few 10$^7$~year for
matter to flow from the core to the tips of the secondary lobes.
This is less than
the radiative age, $\sim$10$^8$~year, of $X$-shaped sources assuming
equipartition of energy between the radiating particles and magnetic field
and 610~MHz as the break frequency of the injected electron population,
and/or the typical radio source lifetime of 10$^8$~year \citep{Rottmann}.
Furthermore, this being an optimistic estimate, the true lobe
lengths are larger if projection effects are taken into account,
which would increase the required flow speeds. On top, once the
backflowing material leaves the primary lobes and has to penetrate
into the ambient medium we expect the flow to be decelerated by
ram-pressure. Therefore, the required flow duration will thus be
closer to a few 10$^8$~year, which would make backflow unlikely
as the dominant formation scenario unless $X$-shaped sources
prove to be unusually~old.

\subsubsection{Buoyancy}

On morphological grounds, the lobes of a radio galaxy have a
lower density than the surrounding medium (Williams 1991).
Therefore, it is expected that buoyancy may have an impact on
the large scale morphology of the radio lobes. \citet{Worralletal}
have applied such a buoyancy model to NGC~326 and is unable to
account for the formation of $X$-shaped radio galaxies.
X-ray studies of gaseous environment of several such sources
were not able to detect significant cluster gas emission \citep{Kraftetal} with
the exception of NGC~326.  Although, this might be due to lack
of sensitivity, the buoyancy model is also faced with the problem
of rotational symmetry of $X$-shaped radio galaxies. The two
angles between the primary and secondary lobes of an $X$-shaped
source are typically equal to within $\pm$10$^\circ$, and
if buoyancy would be a dominant formation process one would expect to find
a more random distribution of these angles \citep{Rottmann}.
Also, the radio source must do significant work on the medium,
and the total work must be comparable to the stored energy within the
lobes \citep[][1999]{kaiser97}.
Typically, the time-scale for formation of such sources,
or the corresponding radius out to which the source expanded,
is the age of the source times the average Mach number ($<$ 1) for expansion
(Alexander 2002), which is an order of magnitude less than age of the source.
In addition, the swept-up gas would
become Rayleigh-Taylor unstable since the density of the cocoon
is very much less than the external swept-up material \citep{scheuer}.

Therefore, we also conclude that buoyancy will influence the large scale
structure of radio galaxies only in dense cluster environments,
as is seen for wide-angle-tail radio sources in cluster of galaxies,
and it is implausible that buoyancy, without a favourable configuration
of the interstellar medium or intergalactic medium, would influence the
structure of $X$-shaped sources.

\subsubsection{Conical precision}

The conical precession model \citep{Parmaetal,Macketal}
requires not only a fortuitous angle between
the precession cone and angle to the line of sight,
but also a happy accident of the positions at which
the source first switched on and its position now.
It therefore seems unlikely that we can explain the number
of such sources seen, or the lack of other related sources
\citep{LeahyParma}.

More importantly, the morphologies of these sources do not seem to fit
conical precision model \citep{Dennett}.
Structures linking the wing and the lobe
should at least be detected at low radio frequencies,
if the morphology arose from a special projection of a
slowly precessing source. Instead, in almost all the sources,
there is a notable lack of such a feature.
Furthermore, the wings being embedded well into the base
also argues against any interpretation in terms of slow motion
of the jet axis. Given these arguments, it is unimaginable that
such precession can be used as an explanation for $X$-shaped sources.

\subsubsection{Reorientation of the jet axis}

Although, at this point, we conclude that the most likely
formation process of $X$-shaped radio galaxies is reorientation of
their jet axis due to a minor merger, an apparent contradiction of
the model is posed by the small number of $X$-shaped radio galaxies
as compared to the rather large number of minor merger events.
The typical duration of the AGN phase is $\sim$10$^7$~yr.
Since, time scale for jet reorientation is short, $\lesssim$ 10$^7$~yr
\citep{Dennett}, or it occurs instantaneously \citep{MerrittEkers},
\citet{MerrittEkers} predict both types of source, $X$-shaped galaxy and
radio galaxy, to be visible for a similar time.
In other words, similar number of sources for both types are expected.

Certainly, selection effects due to projection and beaming can
conceal the $X$-shaped nature of some sources on unfavourable viewing
angles. For moderate intrinsic angles the fraction of sources
hidden by selection effects is $\sim$25\% \citep{Rottmann}.
This suggests that, unless there is a large, hidden population
of objects with very small reorientation angles, selection effects
are not significant and insufficient to account for small number of
$X$-shaped sources.  Another problem for the connection of merging
and reorientation arises when inspecting the environments in which
the hosts of $X$-shaped sources are embedded. None of the known
sources lay in dense clusters, only a few sources seem to be located
in small, poor clusters or groups and rest of the sources seem to
be isolated field galaxies. Although, the latter problem, could be
reconciled in a minor merger model, leading to a sudden flip in the direction
of any associated jet \citep{Merritt}, we still need to address
the small number of $X$-shaped radio galaxies as compared to the
rather large number of minor merger events.

\subsection{Implications on the formation models}
\label{implications}

We now discuss the implications on the assumptions of the
spectral ageing method and discuss what could be the realistic
formation model of $X$-shaped sources.

\subsubsection{The assumptions of the spectral ageing method}

It is possible that the assumptions used in the spectral ageing method
for estimating the age needs to be investigated.
Due to the presence of all category of sources, A, B \& C,
one of the assumptions, the low-surface-brightness wings are in the process
of becoming new active jets, mentioned earlier \citep{LalRao2005}
does not seem true. 
Furthermore, presence of exotic reacceleration mechanism is unlikely
because of absence of any shock signatures. Also, the high
degree of polarised emission observed in the wings \citep{Rottmann}
indicates that the stochastic reacceleration by plasma turbulence
is not applicable in the wings of these sources. 
A combination of rest of the two assumptions, {\it i.e.}
the injection spectral index is varying \citep{Palmaetal},
and presence of a gradient in magnetic field across the source,
together with curved electron energy spectrum \citep {Blundell}
could explain each of these $X$-shaped sources individually,
but a single model presently seems implausible.

\subsubsection{Environments of $X$-shaped sources}

It has also been argued that the $X$-shaped morphology is essentially
a hydrodynamic phenomenon which is a result of supersonic or
buoyant flow of radio plasma in an asymmetric gas distribution.
\citet{LeahyWilliams} and \citet{Worralletal} argues that the
$X$-shaped radio morphology is a result of strong backflow of
material behind the terminal hot spots
of radio galaxy jets and subsequent buoyant evolution of the wings.
In this model, the lobes and the wings must have been supersonic
at some time in the past, but are now evolving buoyantly.
\citet{Capettietal} hypothesized that the $X$-shaped morphology
is a direct result of the supersonic expansion and/or inflation of
the lobe into an elliptical atmosphere and that all radio galaxies
in such environments should exhibit this phenomenon.  In this scenario,
both the lobes and the wings should be enormously overpressurized
relative to the ambient medium \citep{Kraftetal}.
Although inhomogeneous and incomplete, the sample of $X$-shaped sources
show diverse spectral morphologies and a single model with similar
dynamics of the backflow/buoyancy, jet velocities,
density contrast between the jet and the ambient medium,
temperature and density profile of the medium, and
morphology of ambient medium
for these sources would be a challenge.

\subsubsection{Existing formation models {\it versus} Merger of two AGNs}

\citet{LalRao2005} proposed to use the low frequency spectra
at different locations in the source, to distinguish between the
formation models for these sources.
In the simplest picture, the low-surface-brightness wings would have an
older population of the electrons and therefore should have
steeper spectral index as compared to the
active high-surface-brightness radio lobes.
But, the two categories of sources, namely, sources showing the wings
to be of flatter spectral indices as compared to the active lobes
(category A) and sources showing the wings and the active lobes
to be of comparable spectral indices (category B),
listed above does not support this simple picture.
Instead, the third category of sources, five of the twelve sources,
supports this picture.  Furthermore, none of the models mentioned earlier,
support the radio results from the first two categories of sources.
On top, each of these models have their own limitations,
which are independent of our spectral results.

\citet{Begelman1980} first suggested the possibility that AGN
might contain massive binary black hole. The proposition is justified
via., (i) the nuclei of most galaxies contain massive black hole
and (ii) galaxies often merge.  More importantly, all radio images
of the sample of $X$-shaped sources show the wings and the lobes
to be embedded well into the base and argues in favour of 
unresolved, coalescing binary AGN system.
In order to understand, if the $X$-shaped sources are indeed examples
of unresolved binary AGN systems, with two pairs of jets associated
with two unresolved AGNs, several methods have been suggested, {\it e.g.}
an indirect hint for the presence of binary AGNs is using {\it HST} to
identify inwardly decreasing surface brightness profiles in the galaxy.
We have re-analyzed the archived snapshot {\it HST} data for 3C~52,
3C~136.1, 3C~223.1, 3C~315, 3C~403 and 3C~433. A
close inspection of the deconvolved brightness profiles in each source
does not suggest a centrally depressed, nearly flat core.
But is this due to obscuration of the core by the dusty disk
\citep{deKoff}, needs to be followed with new deep images.

\subsection{Do $X$-shaped sources form a single class?}
\label{single}

Based on the GMRT results presented above, we conclude that earlier models
do not explain the formation scenario for the $X$-shaped sources.
Only viable model is the `alternative' model,
{\it i.e.} the $X$-shaped sources consists of two pairs of jets,
which are associated with two unresolved AGNs \citep{LalRao2005}.
This `alternative' model not only explains earlier observational results,
it also explains our low frequency spectral results.
The proposition also supports the small number of $X$-shaped sources,
as the frequency of merger of two field AGNs is small and is definitely
smaller than the number of minor merger events.  Hence, we believe that
every $X$-shaped source consist of an unresolved binary AGN,
giving two pairs of jets corresponding to two AGNs.
Whether or not the central binary black hole can actually
account for the formation of $X$-shaped radio galaxies mainly
depends on the timescales of shrinking of separation radii and
final merging of the binary system. In other words,
it critically depends on the evolution and the stability
of the binary system \citep{Begelman1980}.

We suggest high resolution, multi-frequency, phase-referenced
very-long-baseline-interferometry (VLBI) imaging of
$X$-shaped sources, in order to determine the recent active jet
and investigate if these sources are example of resolved binary AGN systems
\citep{Sudouetal,Rodriguez}.
In addition, we also suggest deep {\it HST}, {\it Chandra} images
to detect binary supermassive black holes, and
a search process using the images from the
VLA NVSS \citep{Condonetal} and FIRST \citep{Beckeretal} surveys with
a goal of increasing the total number of $X$-shaped radio sources,
thereby forming a homogeneous complete sample.
Additional observational results and wisdom gained from it
would allow us to understand these sources in a statistical manner.

\section{Conclusions}
\label{conclusions}

We have presented the lowest frequency images of the sample
of $X$-shaped sources at 240 and 610~MHz,
and our radio spectral results have been instrumental
in testing the formation scenario of these sources.
The measurements presented here
represent most of the database that we require for rigorously testing
and understanding the formation models of these sources.
Based on our careful analysis and
estimation of the possible systematic errors, along with
the integrated radio flux densities and the spectral indices from it
at several locations across each of these sources,
we show that these sources divide into following three categories:

\begin{enumerate}
\item[(A)] Sources in which the wings are of flatter spectral indices
than the active lobes, namely, 4C~12.03, 3C~223.1 and B1059$+$169.
\item[(B)] Sources in which the wings and the active lobes have comparable
spectral indices, namely, 3C~192, B2~0828$+$32, 4C~48.29 and 3C~403.
\item[(C)] Sources in which the wings are of steeper spectral indices
than the active lobes, namely, NGC~326, 3C~52, 3C~136.1, 3C~315
and 3C~433.
\end{enumerate}

These GMRT results do not support earlier known formation models
for the $X$-shaped sources.
While it is equally probable that the three
categories, A, B \& C, of sources are unrelated to one another,
a single model to explain these sources is a challenge.
Currently, only possible model is our `alternative' model,
{\it i.e.} the $X$-shaped sources consists of two pairs of jets,
which are associated with two unresolved AGNs \citep{LalRao2005}.

There is definitely a need to understand the proposed formation scenario
for $X$-shaped sources and hence, follow-up work is necessary.
Future, VLBI results, together with deep {\it HST}, {\it Chandra} images
and results from a larger homogeneous complete sample of $X$-shape
sources would be useful in constraining any possible formation model.

\section*{Acknowledgments}

We thank the staff of the GMRT who have made these observations
possible. GMRT is run by the National Centre for Radio Astrophysics
of the Tata Institute of Fundamental Research.
We also thank the anonymous referee for his/her prompt review
of the manuscript and for comments that
lead to improvement of the paper.
DVL thanks R~Nityananda and M~Hardcastle for discussions
and several useful comments.
This research has made use of the NASA/IPAC Extragalactic Database,
which is operated by the Jet Propulsion Laboratory,
Caltech, under contract with the NASA, and
NASA's Astrophysics Data System.


\label{lastpage}


\begin{thebibliography}{99}
\bibitem[\protect\citeauthoryear{Alexander}{2002}] {alexander}
         Alexander, P., 2002, MNRAS, 335, 610
\bibitem[\protect\citeauthoryear{Alexander \& Leahy}{1987}] {AlexanderLeahy1987}
         Alexander, P., Leahy, J.P., 1987, MNRAS, 225, 1
\bibitem[\protect\citeauthoryear{Baars et~al.}{1977}] {Baarsetal}
         Baars, J.W.M., Genzel, R., Pauliny-Toth, I.I.K., Witzel, A.,
         1977, A\&A, 61, 99
\bibitem[\protect\citeauthoryear{Becker, White \& Edwards}{1991}] {BeckerWhite}
         Becker, R.H., White, R.L., Edwards, A.L., 1991, ApJS, 75, 1
\bibitem[\protect\citeauthoryear{Becker, White \& Helfand}{1995}] {Beckeretal}
         Becker, R.H., White, R.L., Helfand, D.J., 1995, ApJ, 450, 559
\bibitem[\protect\citeauthoryear{Begelman, Blandford \& Rees}{1980}] {Begelman1980}
         Begelman, M.C., Blandford, R.D., Rees, M.J., 1980, Nature, 287, 307
\bibitem[\protect\citeauthoryear{Blundell \& Rawlings}{2000}] {Blundell}
         Blundell, K.M., Rawlings, S., 2000, AJ, 119, 1111
\bibitem[\protect\citeauthoryear{Black et~al.}{1992}] {Blacketal}
         Black, A.R.S., Baum, S.A., Leahy, J.P., Perley, R.A., Riley, J.M.,
         Scheuer, P.A.G., 1992, MNRAS, 256, 186
\bibitem[\protect\citeauthoryear{van~Breugel et~al.}{1983}] {vanBreugeletal}
	van~Breugel, W., Helfand, D., Balick, B., Heckman, T., Miley, G.,
        1983, AJ, 88, 40
\bibitem[\protect\citeauthoryear{van~Breugel \& J\"agers}{1982}]
                                   {vanBreugelJagers}
         van~Breugel, W., J\"agers, W., 1982, A\&AS, 49, 529
\bibitem[\protect\citeauthoryear{Burns, Gregory \& Holman}{1981}] {Burns1981}
         Burns, J.O, Gregory, S.A., Holman, G.D., 1981, ApJ, 250, 450
\bibitem[\protect\citeauthoryear{Capetti et~al.}{2002}] {Capettietal}
         Capetti, A., Zamfir, S., Rossi, P., Bodo, G., Zanni, C.,
         Massaglia, S., 2002, A\&A, 394, 39
\bibitem[\protect\citeauthoryear{Condon et~al.}{1998}] {Condonetal}
         Condon, J.J., Cotton, W.D., Greisen, E.W., Yin, Q.F., Perley, R.A.,
         Taylor, G.B., Broderick, J.J., 1998, AJ, 115, 1693
\bibitem[\protect\citeauthoryear{Dennett-Thorpe et~al.}{1999}] {Dennett1}
         Dennett-Thorpe, J., Bridle, A.H., Laing, R.A., Scheuer, P.A.G.,
         1999, MNRAS, 304, 271
\bibitem[\protect\citeauthoryear{Dennett-Thorpe et~al.}{2002}] {Dennett}
         Dennett-Thorpe, J., Scheuer, P.A.G., Laing, R.A., Bridle, A.H.,
         Pooley, G.G., Reich, W., 2002, MNRAS, 330, 609
\bibitem[\protect\citeauthoryear{Ekers et~al.}{1978}] {Ekers1978}
         Ekers, R.D., Fanti, R., Lari, C., Parma, P., 1978, Nature, 276, 588
\bibitem[\protect\citeauthoryear{Fanaroff \& Riley}{1974}] {FanaroffRiley}
         Fanaroff, B.L., Riley J.M., 1974, MNRAS, 167, 31P
\bibitem[\protect\citeauthoryear{Feretti et~al.}{1983}] {Ferettietal}
         Feretti, L., Giovannini, G., Gregorini, L., Parma, P.,
         1983, A\&A, 126, 311
\bibitem[\protect\citeauthoryear{Ficarra, Grueff \& Tomassetti}{1985}]
                                {Ficarraetal}
         Ficarra, A., Grueff, G., Tomassetti, G., 1985, A\&AS, 59, 255
\bibitem[\protect\citeauthoryear{Gregory \& Condon}{1991}]{GregoryCondon1991}
         Gregory, P.C., Condon, J.J., 1991, ApJS, 75, 1011
\bibitem[\protect\citeauthoryear{Gopal-Krishna, Biermann \& Wiita}{2003}]
                                {Krishna}
         Gopal-Krishna, Biermann, P.L., Wiita, P.J., 2003, ApJL, 594, 103
\bibitem[\protect\citeauthoryear{Gower et~al.}{1967}]{Goweretal1967}
         Gower, J.F.R., Scott, P.F., \& Wills, D. 1967, Mem.RAS, 71, 49
\bibitem[\protect\citeauthoryear{Hardcastle et~al.}{1997}] {Hardcastleetal}
	 Hardcastle, M.J., Alexander, P., Pooley, G.G., Riley, J.M.,
	 1997, MNRAS, 288, 859
\bibitem[\protect\citeauthoryear{Heckman et~al.}{1994}] {Heckman1994}
         Heckman, T.M., O'Dea, C.,P., Baum, S.A., Laurikainen, E.,
         1994, ApJ, 428, 65
\bibitem[\protect\citeauthoryear{H\"ogbom}{1979}] {Hogbom}
         H\"ogbom, J.A., 1979, A\&AS, 36, 173
\bibitem[\protect\citeauthoryear{Kaiser \& Alexander}{1997}] {kaiser97}
         Kaiser, C.R., Alexander, P., 1997, MNRAS, 286, 215
\bibitem[\protect\citeauthoryear{Kaiser \& Alexander}{1999}] {kaiser99}
         Kaiser, C.R., Alexander, P., 1999, MNRAS, 305, 707
\bibitem[\protect\citeauthoryear{Kellermann}{1969}] {Kellermann}
         Kellermann K.I., Pauliny-Toth, I.I.K., Williams, P.J.S,
         1969, ApJ, 157, 1
\bibitem[\protect\citeauthoryear{de~Koff}{1996}] {deKoff}
         de Koff, S., Baum, S.A., Sparks, W.B., Biretta, J., Golombek, D.,
         Macchetto, F., McCarthy, P., Miley, G.K., 1996, ApJS, 107, 621
\bibitem[\protect\citeauthoryear{Kraft et~al.}{2005}] {Kraftetal}
         Kraft, R.P., Hardcastle, M.J., Worrall, D.M., Murray, S.S.,
         2005, ApJ, 622, 149
\bibitem[\protect\citeauthoryear{Kuhr et~al.}{1981}] {Kuhretal1981}
         Kuhr, H., Witzel, A., Pauliny-Toth, I.I.K., Nauber, U.
         1981, A\&AS, 45, 367
\bibitem[\protect\citeauthoryear{Laing, Riley \& Longair}{1983}] {Laing1983}
         Laing, R.A., Riley, J.M., Longair, M.S., 1983, MNRAS, 204, 151
\bibitem[\protect\citeauthoryear{Lal \& Rao}{2005}] {LalRao2005}
         Lal, D.V., Rao, A.P., 2005, MNRAS, 356, 232
\bibitem[\protect\citeauthoryear{Large et~al.}{1981}] {Largeetal1981}
         Large, M.I., Mills, B.Y., Little, A.G., Crawford, D.F., Sutton, J.M.,
         1981, MNRAS, 194, 693
\bibitem[\protect\citeauthoryear{Leahy et~al.}{1997}] {Leahyetal97}
         Leahy, J.P., Black, A.R.S., Denett-Thorpe, J., Hardcastle, M.J.,
         Komissarov, S., Perley, R.A., Riley, J.M., Scheuer, P.A.G.,
         1997, MNRAS, 291, 20
\bibitem[\protect\citeauthoryear{Leahy \& Parma}{1992}] {LeahyParma}
         Leahy, J.P., Parma, P., 1992
         in Roland, J., Sol, H., Pelletier, G, eds,
         Extragalactic Radio Sources.
         From Beams to Jets, Cambridge University Press, p.~307
\bibitem[\protect\citeauthoryear{Leahy \& Williams}{1984}] {LeahyWilliams}
         Leahy, J.P., Williams, A.G., 1984, MNRAS, 210, 929
\bibitem[\protect\citeauthoryear{Mack et al.}{1994}] {Macketal}
         Mack, K.-H., Gregorini, L., Parma, P., Klein, U.,
         1994, A\&AS, 103, 157
\bibitem[\protect\citeauthoryear{Martel et~al.}{1999}] {Martel1999}
         Martel, A.R., Baum, S.A., Sparks, W.B., Wyckoff, E., Biretta, J.A.,
         Golombek, D., Macchetto, F.D., de~Koff, S., McCarthy, P.J, Miley, G.K.,
         1999, ApJS, 122, 25
\bibitem[\protect\citeauthoryear{McCarthy, Spinard \& van~Breugel}{1995}]
                                {McCarthy1995}
         McCrathy, P., Spinard, H., van Breugel, W., 1995, ApJS, 99, 27
\bibitem[\protect\citeauthoryear{Merritt}{2004}] {Merritt}
         Merritt, D., 2004, in Coevolution of black holes and Galaxies,
         Ho., L.C., Ed. (Cambridge Univ Press)
\bibitem[\protect\citeauthoryear{Merritt \& Ekers}{2002}] {MerrittEkers}
         Merritt, D., Ekers, R.D., 2002, Sci, 297, 1310
\bibitem[\protect\citeauthoryear{Murgia et~al.}{2001}] {Murgiaetal}
         Murgia, M., Parma, P., de~Ruiter, H.R., Bondi, M., Ekers, R.D.,
         Fanti, R., Fomalont, E.B., 2001, A\&A, 380, 102
\bibitem[\protect\citeauthoryear{Owen \& Ledlow}{1997}] {OwenLedlow}
         Owen, F.N., Ledlow, M.J., 1997, ApJS, 108, 41
\bibitem[\protect\citeauthoryear{Palma et al.}{2000}] {Palmaetal}
         Palma, C., Bauer, F.E., Cotton, W.D., Bridle, A.H., Majewski, S.R.,
         Sarazin, C.L., 2000, AJ, 119, 2068
\bibitem[\protect\citeauthoryear{Parma, Ekers \& Fanti}{1985}] {Parmaetal}
         Parma, P., Ekers, R.D., Fanti, R., 1985, A\&AS, 59, 511
\bibitem[\protect\citeauthoryear{Pilkington \& Scott}{1965}]{PilkingtonScott1965}
         Pilkington, J.D.H., Scott, P.F., \& Wills, D. 1965, Mem.RAS, 69, 183
\bibitem[\protect\citeauthoryear{Rees}{1978}] {Rees}
         Rees, M.J., 1978, Nat, 275, 516
\bibitem[\protect\citeauthoryear{Rottmann}{2001}] {Rottmann}
         Rottmann, H., 2001, PhD thesis,
         Rheinischen Friedrich-Wilhelms-Univesit\"at Bonn
\bibitem[\protect\citeauthoryear{Rodriguez et~al.}{2006}] {Rodriguez}
         Rodriguez, C., Taylor, G.B., Zavala, R.T., Peck, A.B.,
         Pollack, L.K., Romani, R.W. 2006, ApJ, 646, 49
\bibitem[\protect\citeauthoryear{Ryle}{1960}] {Ryle}
         Ryle, M., 1960, J. Instn elect. Engrs, 6, 14
\bibitem[\protect\citeauthoryear{Sandage}{1972}] {Sandage1972}
         Sandage, A., 1972, ApJ, 178, 25
\bibitem[\protect\citeauthoryear{Scheuer}{1974}] {scheuer}
         Scheuer, P.A.G., 1974, Sci, 300, 1263
\bibitem[\protect\citeauthoryear{Scheuer}{1995}] {Scheuer95}
         Scheuer, P.A.G., 1995, MNRAS, 277, 331
\bibitem[\protect\citeauthoryear{Steer, Dewdney \& Ito}{2003}] {Steeretal}
         Steer, D., Dewdney, P., Ito, M.,
         2003, Sci, 300, 1263
\bibitem[\protect\citeauthoryear{Sudou, Iguchi \& Murata}{2003}] {Sudouetal}
         Sudou, H., Iguchi, S., Murata, Y., 2003, Sci, 300, 1263
\bibitem[\protect\citeauthoryear{Swarup et al.}{1991}] {Swarupetal}
         Swarup, G., Ananthakrishnan, S., Kapahi, V.K., Rao, A.P.,
         Subrahmanya, C.R., Kulkarni, V.K., 1991, Cu. Sc., 60, 95
         2003, ApJ 126, 113
\bibitem[\protect\citeauthoryear{Ulrich \& R\"onnback}{1996}] {UlrichRonnback}
         Ulrich, Marie-Helene, R\"onnback, Jari, 1996, A\&A, 313, 750
\bibitem[\protect\citeauthoryear{Wirth, Smarr \& Gallagher}{1982}] {Wirth1982}
         Wirth, A., Smarr, L., Gallagher, J.S., 1982, AJ, 87, 602
\bibitem[\protect\citeauthoryear{White \& Becker}{1992}] {WhiteBecker}
         White, R.L., Becker, R.H., 1992, ApJS, 79, 331
\bibitem[\protect\citeauthoryear{Worrall, Birkinshaw \& Cameron}{1995}] {Worralletal}
         Worrall, D.M., Birkinshaw, M., Cameron, R.A., 1995, ApJ, 449, 93
\bibitem[\protect\citeauthoryear{Zirbel}{1997}] {Zirbel1997}
         Zirbel, E.L., 1997, ApJ, 476, 489
\bibitem[\protect\citeauthoryear{Zwicky \& Kowal}{1968}] {ZwickyKowal1968}
         Zwicky, F., Kowal, C.T., 1968,
         Catalogue of Galaxies and Clusters of galaxies, Pasadena, Caltech

\end{thebibliography}
\end{document}